\documentclass[review]{elsarticle}

\usepackage{lineno,hyperref}
\usepackage{booktabs,color}
\usepackage{amssymb}
\usepackage{threeparttable}
\modulolinenumbers[5]
\usepackage{framed}
\usepackage{makecell}
\usepackage{multirow} 
\usepackage{graphicx} % Required for inserting images

\begin{document}

%\maketitle

\begin{frontmatter}

\title{Can Adjusting Hyperparameters Lead to Green Deep Learning: An Empirical Study on Correlations between Hyperparameters and Energy Consumption of Deep Learning Models}

\author[mymainaddress]{Taoran Wang}
\author[mymainaddress]{Yanhui Li}
\author[mymainaddress]{Mingliang Ma}
\author[mymainaddress]{Lin Chen}
%\author[secondaddress]{Wenjun Ke}
\author[mymainaddress]{Yuming Zhou}

\address[mymainaddress]{State Key Laboratory for Novel Software Technology, Nanjing University, 210023, China}
%\address[secondaddress]{School of Computer Science and Engineering, Southeast University, Nanjing, 211189, China}
%\address[mysecondaryaddress]{Department of Computer Science and Technology, Nanjing University, 210023, China}
%\address[mythirdaddress]{}

\begin{abstract}
%This template helps you to create a properly formatted \LaTeX\ manuscript.
%Background: 

\noindent \textbf{Context:} Along with developing Deep learning (DL) models, larger datasets and more complex model structures are applied, leading to rising computing resources and energy consumption, which is an alert that green DL models should receive more attention.

%Deep learning (DL) gains a lot of attention in software engineering. However, many researchers only focus on the performance of the models instead of the energy consumption. Declining the energy consumption of DL software can not only reduce carbon dioxide production but can also reduce the financial cost.

\noindent \textbf{Objective:} This paper focuses on a novel view to analyze DL energy consumption: the effect of hyperparameters on the energy cost of DL models.

\noindent \textbf{Method:} Our approach involves using mutation operators to simulate how practitioners adjust hyperparameters, such as epochs and learning rates. We train the original and mutated models separately and gather energy information and run-time performance metrics. Moreover, we focus on the parallel scenario where multiple DL models are trained in parallel.

\noindent \textbf{Results:} To examine the effect of hyperparameters on energy consumption, we conducted extensive experiments on five real-world DL models. The results show that (1) many hyperparameters studied have a (positive or negative) correlation with energy consumption, (2) adjusting hyperparameters can make DL models greener, i.e., lead to less energy consumption without performance damage, and (3) in a parallel environment, energy consumption becomes more susceptible to change.

\noindent \textbf{Conclusions:} We suggest that hyperparameters need more attention in developing DL models, as appropriately adjusting hyperparameters would cause green DL models.

\end{abstract}

\begin{keyword}
Green software \sep energy consumption \sep deep learning \sep mutation testing \sep run-time performance 
\end{keyword}

\end{frontmatter}

\section{Introduction}
With the enhancement of computational power~\cite{mittal2019survey} and the support of big data~\cite{schuhmann2022laion}, DL technology has made breakthrough advancements in the past few decades~\cite{lecun2015deep}. It has achieved an extremely high level of intelligence in many fields, such as natural language processing~\cite{otter2020survey}, image recognition~\cite{chen2019looks}, and medical diagnosis~\cite{liu2022deep}. Models based on DL technology have approached or even surpassed the level of human experts~\cite{silver2017mastering,lan2022alphazero}. However, as DL technology follows a data-driven paradigm~\cite{zhang2020machine}, the noise and bias~\cite{labib2022analysis} in the training data inevitably lead to errors in the DL models trained~\cite{islam2019comprehensive}, creating unexpected outputs in practical applications and resulting in performance loss~\cite{zhao2021state}. The frequent occurrence of these errors\footnote{\url{https://en.wikipedia.org/wiki/List_of_self-driving_car_fatalities}} raises concerns about the reliability of DL models, prompting researchers to focus on their quality assurance~\cite{ma2018secure,shen2020multiple,felderer2021quality}.

When we view DL models as a new type of software, introducing software engineering (SE) methods and ideas is a significant way to achieve DL quality assurance~\cite{DBLP:journals/software/Menzies20}. Existing research has shown that software developers can reduce the abnormal behavior of models and subsequently improve their quality by dealing with quality issues within DL models~\cite{DBLP:journals/corr/abs-1905-05786}. Besides proposing specific SE methods for DL models, such as analysis and testing~\cite{DBLP:conf/sosp/PeiCYJ17,DBLP:conf/issta/WangLHCZZ23,zhang2021cagfuzz}, to directly improve the quality of models, many SE-related studies focus on empirical research~\cite{DBLP:conf/issre/0001GMLK19} on the life cycle of DL models. In terms of \textit{model performance}, 
Yan et al.~\cite{DBLP:conf/sigsoft/YanTLZMX020} investigated the correlation between the quality of DL models and coverage criteria and observed that they are not correlated. Hu et al.~\cite{DBLP:journals/tosem/HuGCXMPT22} executed an extensive empirical investigation to shed light on how the retraining process and data distribution influence the performance of DL models. Regarding \textit{model fairness}, Biswas and Rajan~\cite{DBLP:conf/sigsoft/BiswasR20} concentrated on empirically evaluating the fairness and mitigations of 40 top-rated Kaggle models with five tasks. They found that some model optimization techniques may increase the unfairness of studied models. Gohar et al.~\cite{DBLP:conf/icse/GoharBR23} meticulously investigate ensemble techniques (including Bagging, Boosting, Stacking, and Voting) in real-world applications with an extensive benchmark of 168 ensemble models. They pointed out that designing fairer ensembles is possible without mitigation techniques. 

Recently, researchers have been paying attention to the \textit{energy consumption} of DL models. Along with developing DL models, larger datasets and more complex model structures are applied, which leads to more computing resources and rising energy consumption. According to the observation \cite{BLEU-a-method} of Papineni et al., the state-of-the-art DL models cost $3,000,000\times$ resources from 2012 to 2018. The high energy consumption produces more carbon dioxide \cite{tong2020economic}, which harms our environment and also increases the financial cost of DL model development and maintenance. This extreme increment in energy consumption of DL models is an alert that green DL models should receive more attention. To our knowledge, the SE studies on the energy consumption of DL models are still in the early stages. Georgiou et al.~\cite{greenAi} is the first empirical research on green DL software, which conducts an in-depth analysis of DL frameworks and shows the difference between DL frameworks in energy consumption.

Inspired by Georgiou et al.'s study~\cite{greenAi}, this paper focuses on another view to analyze DL energy consumption: the effect of hyperparameters on the energy cost of DL models. Our literature review shows that although DL hyperparameters are known to influence model performance~\cite{akiba2019optuna,hertel2020sherpa,li2023survey} (e.g., accuracy), the correlation between hyperparameters and energy consumption is still not revealed. Specifically, this paper aims to answer the following questions: ``How can hyperparameters affect energy consumption?'' and ``Can we employ such affection to make DL models greener?''

%This paper deviates from previous studies. For the energy consumption factors, we focus on DL hyperparameters, which are known to have influences on accuracy. However, the correlation between hyperparameters and energy consumption is still not revealed. Therefore, this paper aims to figure out whether hyperparameters can affect energy consumption and how hyperparameters affect energy consumption. Besides, if adjusting hyperparameters can change the energy consumption, can it make the models greener? 

 Our approach comprises three steps to investigate the correlations between hyperparameters and energy consumption: (a) we select open-source DL based software and employ mutation operators~\cite{DeepFD}\footnote{During the mutation testing, mutation operators are usually designed to seed faults and check whether test cases could identify the seeded faults. Here, we apply this technique to generate mutated models with different hyperparameter settings.} to imitate how DL practitioners adjust hyperparameters, e.g., epochs and learning rates (see details in Section~\ref{sec:mutation}); (b) We train the original and mutated DL models separately and collect energy information with run-time performance metrics (e.g., package, RAM, and GPU). In addition, we focus on the \textit{parallel scenario}, in which DL models are trained in parallel and metrics are collected in parallel environments (see Section~\ref{sec:collection}); (c) we conduct several analysis methods to obtain our conclusion, including Spearman correlation analysis, trade-off analysis, and parallel analysis (see Section~\ref{sec:analysis}).

% For the mutation test, instead of seeding faults, we apply this technique to mimic how DL practitioners adjust hyperparameters. In previous research, the range of mutation is larger than in common usage when seeding faults. In this paper, we decrease the range to make it small enough, which is more likely to be the range DL practitioners shall operate. We also find that in practice, multiple DL software is usually run on one server. Therefore, we introduce the parallel environment to discover the distinctions between this and training models in the common environment. 
% Empirical research is completed on five real-world DL based software to investigate the correlations between hyperparameters and energy consumption. These open-source programs are collected from GitHub according to our mutation test framework. Then, we mutate the hyperparameters of the models to imitate how DL practitioners adjust the hyperparameters and collect the metrics of the energy consumption and run-time performance while the models are training. With these data, we conducted a Spearman correlation analysis to answer our questions. Finally, to uncover the correlations in a parallel environment where several models are usually trained on a single server, we run our mutation test in parallel and re-do the analysis above.

In our evaluation, we conducted empirical research on five real-world DL based software~\cite{mnist, mff, siamese, resnet, HRNet-18} and three widely used datasets~\cite{mnist-d, cifar-10, Market}. Our experiments and results are organized into the following research questions (RQs):

\begin{itemize}
	\item \textbf{RQ1: What is the correlation between model energy consumption and their hyperparameters?}
 
	With RQ1, we employ the Spearman correlation analysis to find out how hyperparameters influence the energy consumption and performance of the models. Through the analysis, we observe that \textit{most hyperparameters correlate (positively or negatively) with energy consumption.} 
	%Knowing this information, we can help practitioners and researchers build green models with clearer orientations. 
 
	\item \textbf{RQ2: Can we make DL models greener by adjusting hyperparameters?}
 
	With RQ2, we compare the energy consumption of the mutations with that of the original models in the trade-off analysis. Compared with the original models, \textit{some mutations become greener}, i.e., they consume less energy and achieve comparable or better performance. 
	%With this information, we can prove the feasibility of greener models by adjusting specific hyperparameters.
	
	\item \textbf{RQ3: Are the conclusions above able to apply to the parallel environment?} 
 
	In reality, multiple models train on a single server in parallel nowadays. As a result, we aim to examine the distinction when training models in parallel. 
	For the parallel environment, we discover that \textit{energy consumption is more sensitive to hyperparameters shifting, while performance is more stable.} 
	%From this information, we shall know whether our former results fit the parallel environment and what new results we can find. 
		
\end{itemize}

%We find that most hyperparameters have positive or negative correlations with energy consumption. From our observation, epochs are positively correlated with all hardware energy consumption and run-time performance; the learning rate negatively correlates with RAM energy consumption, GPU energy consumption, and run-time performance; Gamma is negatively correlated with energy consumption. RAM energy consumption is also correlated with weight decay. Compared with the original models, some mutations become greener, which means they cost less energy or gain a better performance. Usually, decreasing epochs in a suitable range can reduce the hardware energy consumption without harming the network performance. Adjusting the learning rate can improve performance with similar energy consumption. Mutating gamma can reduce the package and GPU energy consumption. Adjusting the weight decay can also reduce GPU energy consumption. For the parallel environment, we discover that energy consumption is more sensitive to hyperparameters shifting, while performance is more stable in the mutation test. This means that DL models can easily become greener in the parallel environment.

This paper makes the following contributions.

\begin{enumerate}
    \item\textbf{Approach.} We proposed an energy consumption measurement approach based on DL models mutation. The method based on mutation testing can run mutation operators and collect energy consumption metrics in a hyperparameter space close to the original values. With these metrics, researchers and practitioners can analyze and reveal the relationship between the hyperparameters and energy consumption.
    \item\textbf{Scenario.} We introduced the single/parallel scenario, focusing on DL models trained singly/in parallel. The design of these scenarios was motivated since many models are trained in parallel, such as on a shared server. We evaluated and analyzed the metrics in these scenarios to uncover the differences and similarities between the two scenarios.
    \item\textbf{Study.} We conducted empirical research on six real-world models with their mutations trained singly/in parallel. As for the evaluation, we conducted a correlation analysis, a causal analysis, and an ordinary least square regression to observe the relationships and trade-off issues.
\end{enumerate}

Here is the list of the rest of this paper. Section~\ref{sec:background} explains the relative concepts, including green DL and mutation tests. The mutation operators and analysis methods are presented in Section~\ref{sec:appraoch}. Section~\ref{sec:result} shows the results of the evaluations. Section~\ref{sec:discussion} discusses other details from our observation and insight into our work. Section~\ref{sec:threats} tells the threats to the validity of this work. Section~\ref{sec:related} lists the related work of this paper. The conclusion is made in Section~\ref{sec:conclusion}.

\section{Background}
\label{sec:background}

\subsection{Preliminaries}
This section presents the main aspects of backgrounds: deep learning, energy consumption, and mutation testing and operators.

\subsubsection{Deep learning}

Deep learning (DL) is a subfield of ML with more complex model structures, whose model is a deep neural network with more layers and abstraction~\cite{Deep-Learning}. Due to complex models and sufficient data, DL can study the regular patterns of the data to achieve feature learning and representation learning. 
%DL is now widely used in many domains, including computer vision, natural language processing, speech and audio recognition, and generative models.

Since the increasing usage of DL models, SE research on DL development has steadily grown. Many works have contributed to testing and analyzing criteria for DL systems \cite{DeepGauge-Multi-Granularity, Guiding-Deep-Learning, Structural-Test-Coverage, Apricot-A-Weight, DeepLocalize-Fault-Localization}. Moreover, SE researchers have proposed empirical studies to assess the relationship between various factors in developing DL models and their quality~\cite{greenAi,zhang2020empirical,zhang2019empirical,chen2021empirical}. 

\subsubsection{Energy Consumption}
 The energy consumption is often considered the amount of electrical energy consumed by the hardware \cite{chen2011overview}, usually measured in Joules. 
 Environmental problems are receiving increasing attention, and therefore, more and more researchers are concerned with the carbon footprint of computer programs~\cite{GREEN}. Many studies have proved the importance of reducing the energy consumption of software~\cite{dl-software,cnn-fiancail}. Many previous works mainly focus on embedded systems, but only some notice software and DL models. Recently, researchers paid more attention to the green issues of DL models~\cite{greenAi} and system software~\cite{Twins-or-False}.
 
Given two studied systems (software or models), $\mathcal{A}$ and $\mathcal{B}$, we consider $\mathcal{A}$ greener than $B$ in one of the following two conditions:
 \begin{itemize}
 	\item $\mathcal{A}$ costs less energy than $\mathcal{B}$ but they have similar performance.
 	\item $\mathcal{A}$ consumes energy similar to $\mathcal{B}$, while $\mathcal{A}$ has better performance.
 \end{itemize} 
 A greener system would have many benefits, including less environmental damage and financial cost.

\subsubsection{Mutation Testing and Operators} Mutation testing is a software testing technique that involves designing mutation operators to introduce faults in the software. This technique imitates how programmers make mistakes when writing programs and estimates test cases by checking if they can detect these faults. Mutation testing is widely used in traditional SE studies~\cite{industry}, e.g., López et al.~\cite{java-c} mutated Java and C programs, Álvarez\--{}García et al.~\cite{cpp} updated mutation tool for new C++ standards, and Wen et al.~\cite{api} employed mutation operators to discover API misuse patterns.

With the development of DL models, SE researchers have introduced mutation testing ideas to achieve the quality assurance of DL models. Novel and effective mutation operators have been proposed to test DL models \cite{DeepFD,shen2018munn,ma2018deepmutation,hu2019deepmutation++,humbatova2021deepcrime}. Besides, optimization methods have been pointed out for DL mutation testing to reduce labeling and running expenses \cite{shen2021boundary,li2022higher}. 
%mutation tests on the DL software are more frequently researched. For instance, 35 DL mutation operators are defined by Humbatova et al.~\cite{DeepCrime} according to three empirical studies.
 \textit{Our paper introduces mutation operators to discover the differences and similarities in single/parallel scenarios in the nearby hyperparameter space.}

\subsection{Main idea and motivation example}
Numerous studies exist on modifying the hyperparameters of DL models to enhance performance \cite{hertel2020sherpa,li2023survey}. However, few concentrate on their energy consumption. In this study, we aim to examine whether we can make the training phase of DL models more greener by adjusting hyperparameters.

\begin{figure}[t]
    \centering
    \includegraphics[scale=0.25]{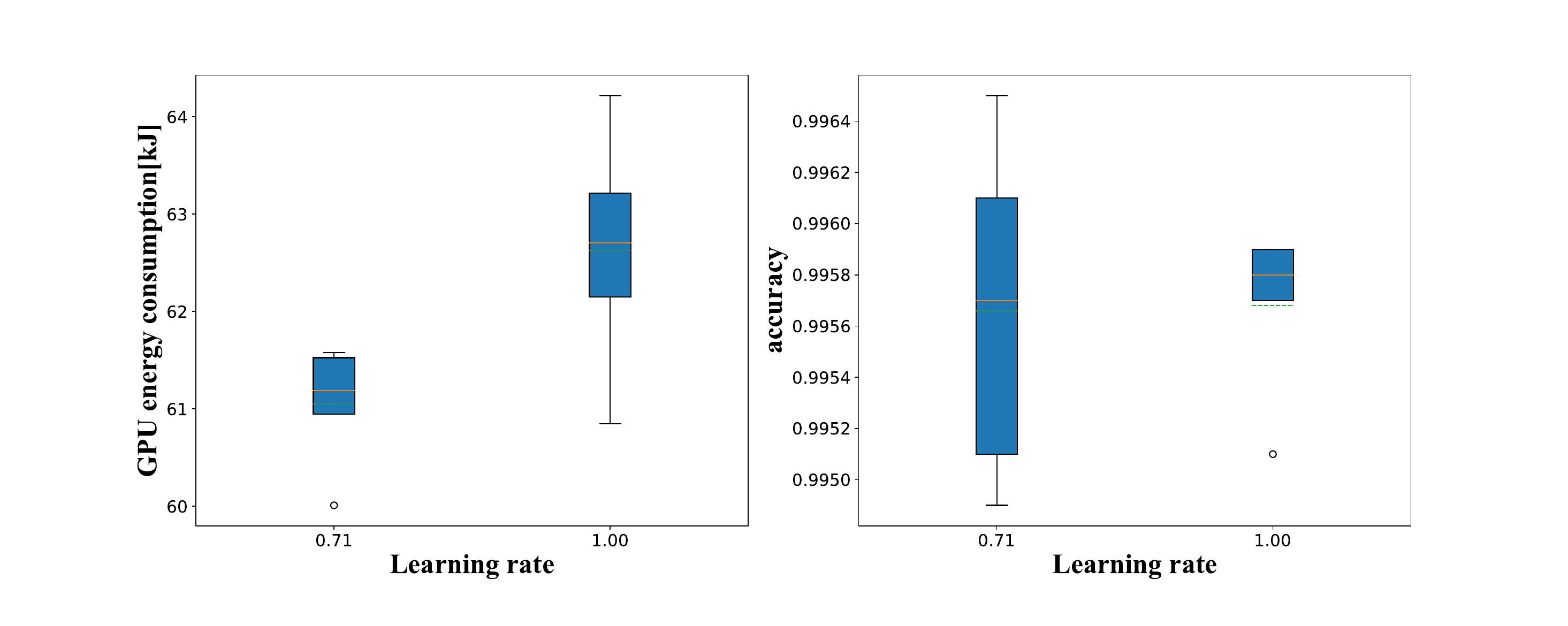}
    \caption{An example of how the learning rate leads to a greener network training phase. The left figure shows the total GPU energy consumption of the training phase. The right figure shows the accuracy of the model tested on the test dataset. The example is a Siamese net base on Resnet20~\cite{resnet18} train on the dataset MNIST~\cite{mnist-d} with different Learning rates. The red line shows the medians, and the green line shows the averages. }
    \label{fig:bg}
\end{figure}

We conducted a primitive investigation on revising learning rates (one of the studied hyperparameters; see details in Table~\ref{tab:op}) of DL models. 
Figure~\ref{fig:bg} shows the difference in GPU energy consumption and accuracy when we adjust the learning rate of the Siamese network. The figure illustrates a decrease in GPU energy consumption corresponding to a reduction in the model's learning rate. The average energy expenditure reduces by about 1.6 kJ, while the average accuracy remains largely unaffected. This example demonstrates that \textit{adjusting hyperparameters can reduce energy consumption during training while maintaining the model's performance}. 

\section{Our approach}
\label{sec:appraoch}
\begin{figure}[t]
    \centering
    \includegraphics[scale=0.32]{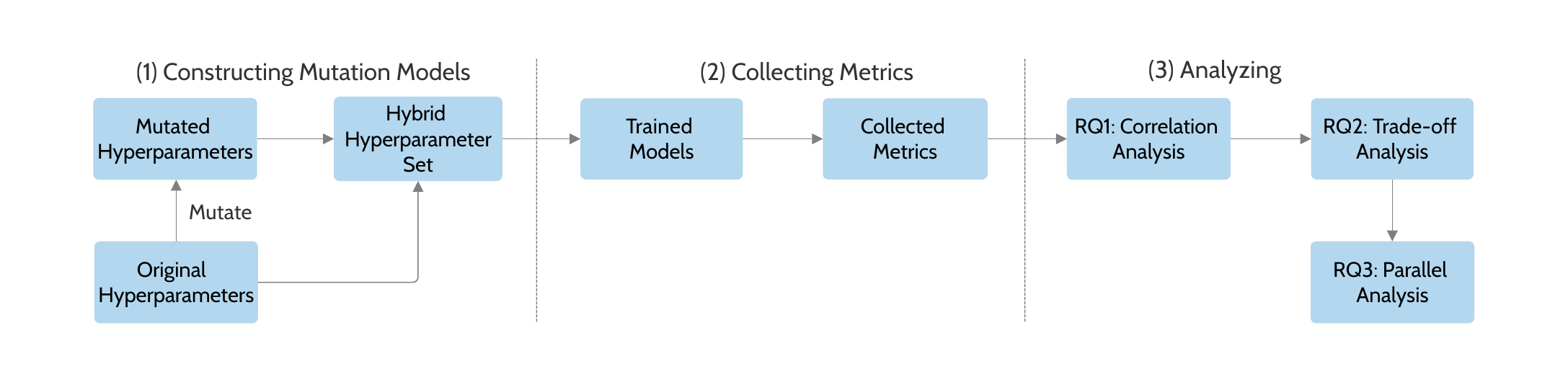}
    \caption{The flowchart of our approach.}
    \label{fig:flow}
\end{figure}

The process flow of our methodology is outlined in Figure~\ref{fig:flow}. We begin by mutating the original hyperparameter settings to generate various models with varying hyperparameters. Then, we employ original and mutated model structures to train the models and collect energy consumption metrics with run-time performance. Ultimately, the collected metric assesses and reports the correlation between hyperparameters and energy consumption.

We conducted our approach on a local server with 20 CPU cores (Intel(R) Xeon(R) Silver 4210R CPU @ 2.40GHz) and GPU (GeForce RTX 3080). The Linux kernel we used is Linux version 5.19.0-45-generic. The main experimental code runs based on Python 3.7.0.

\subsection{Constructing mutated models}
\label{sec:mutation}
\subsubsection{Selecting original models}

Our mutation approach is designed to mutate the models with various hyperparameter settings, train them, and collect metrics to assess energy consumption.
We choose the original DL based software(i.e., DL model structures) with default hyperparameters according to the following rules: 
\begin{itemize}
	\item We can run the codes correctly. According to the code documents, we train the models, test them, compare the results with the documents, and find if there are serious errors and distinctions.
	\item Our approach can mutate the models. Our approach inputs the mutated hyperparameters and runs these codes using command line instructions. Hence, the code should provide these functions to input hyperparameters for command line instructions.
	\item The hyperparameters can be mutated from each model, and it is better to be the same kind. We tried to find models with the same hyperparameters; hence, we can find the correlation for these hyperparameters in different models. %Through this rule, we can reduce the impact of models on energy consumption.
	\item To measure the performance of the mutated training phase, we demand that the models offer indicators to describe the performance of the training, like accuracy. 
\end{itemize}

Five models are chosen, finally listed in Table~\ref{tab:models}, including Models $\mathcal{A}$\footnote{\url{https://github.com/pytorch/examples/tree/main/mnist}}~\cite{mnist}, $\mathcal{B}$\footnote{\url{https://github.com/pytorch/examples/tree/main/mnist\_forward\_forward}} \cite{mff}, $\mathcal{C}$\footnote{\url{https://github.com/pytorch/examples/tree/main/siamese\_network}}~\cite{siamese}, $\mathcal{D}$\footnote{\url{https://github.com/akamaster/pytorch\_resnet\_cifar10}}~\cite{resnet}, and $\mathcal{E}$\footnote{\url{https://github.com/layumi/Person\_reID\_baseline\_pytorch}}~\cite{HRNet-18}. The first two models are Pytotch examples using MNIST~\cite{mnist-d}. Model $\mathcal{C}$ is a Siamese network based on Resnet18~\cite{resnet18}. Model $\mathcal{D}$ is a ResNet20~\cite{resnet18} trained with the dataset CIFAR10~\cite{cifar-10}. Model $\mathcal{E}$ is an HRNet~\cite{hrnet} trained on the dataset Market-1501~\cite{Market}. Note that the last two models have larger datasets (i.e., CIFAR10 and Market-1501) compared to MNIST, and Models $\mathcal{C}$, $\mathcal{D}$, and $\mathcal{E}$ have deeper networks. 

For three studied datasets, MNIST~\cite{mnist-d} consists of handwritten numbers from 250 individuals, of which 50\% are high school students, and 50\% are Census Bureau staff. CIFAR-10~\cite{cifar-10} contains ten categories of RGB color images to identify ubiquitous objects, including 50,000 training images and 10,000 test images in the data set. Market-1501~\cite{Market} has labeled pictures of a total of 1501 pedestrians. Among them, 751 pedestrian annotations are used for the training set, and 750 are used for the testing set.

\begin{table}[t]
    \begin{center}
        \tabcolsep = 1em
        \caption{Details of selected models}
        \label{tab:models}
        \vspace{1ex} 
        \footnotesize
        \resizebox{\linewidth}{!} {
            \begin{tabular}{|c|c|l|}
                \hline
                Dataset &  Model & Structure descriptions \\ \hline\hline
                \multirow{6}*{MNIST} & \multirow{2}*{$\mathcal{A}$} & 
               It is a network consisting of four layers: Conv2d(1, 32, 3, 1), \\ 
                % & \multirow{4}*{It is a simple network with two convolutional layers. } \\
                 ~ & ~ &  Conv2d(32, 64, 3, 1), Linear(9216, 128), and Linear(128, 10). \\

                \cline{2-3}
                ~ & $\mathcal{B}$ & It is a network with two Linear layers. \\
                % & \thead{It a small model using the forward-\\ forward algorithm, which replaces the forward and backward \\ passes of backpropagation.} \\
                \cline{2-3}
                ~ & \multirow{3}*{$\mathcal{C}$} & 
                It is a Siamese network, which is achieved by sharing weights.\\
                 ~ & ~ & It is based on ResNet18~\cite{resnet18}, which is a relatively shallow\\
                 ~ & ~ & ResNet network.\\
                \hline
                \multirow{2}*{CIFAR-10} & \multirow{2}*{$\mathcal{D}$} & 
                It is a special ResNet for CIFAR10 as described in the \\ 
                 ~ & ~ & paper, called ResNet20~\cite{resnet18}.\\
                \hline
                \multirow{2}*{Market-1501} & \multirow{2}*{$\mathcal{E}$} & 
                It uses HRNet-18~\cite{hrnet}, which is a deep network with multiple\\
                 ~ & ~ & branches on human pose estimation problems.\\
                \hline
            \end{tabular}
        }
        \begin{tablenotes}
            \footnotesize
            \item The structure of the first two models is shown in the table. Since the last three models are more complex, this table only lists their model names.
            \end{tablenotes}
    \end{center}
\end{table}

\subsubsection{Mutating hyperparameters}

This study introduces five mutation operators based on the ideas applied in DeepFD~\cite{DeepFD}. The mutation operators in DeepFD are used to generate faults in models. In contrast, our mutation operators change the value of hyperparameters \textit{in a reasonable range}, designed to imitate how DL practitioners adjust the hyperparameters. 

Mutation operators are detailed in Table~\ref{tab:op} with different ranges to randomly generate the mutated values of hyperparameters, where ``d'' means the default value provided by the model authors, and [$x_1$d,$x_2$d] indicates the ranges for mutation. We conducted a preliminary experiment for each mutation operator: if the mutation excessively impacts the model performance, we shall narrow the range; if the mutation of the hyperparameters weakly influences the model performance, we shall use a more extensive range. For example, we observe that mutating epochs in the range between half the default value and two times (i.e., [0.5d, 2d]) the default value has an excessive impact on the accuracy of the models. %The DL practitioners could not polish their model's hyperparameters in this range. 
Therefore, we reduce the mutation range, as shown in Table~\ref{tab:op}. 

\begin{table}[t]
    \begin{center}
        \tabcolsep = 1em
        \caption{The hyperparameters mutated by the operators and the range of each mutation}
        \label{tab:op}
        \vspace{1ex} 
        \small
        \begin{tabular}{|c|c|}
            \hline
            hyperparameter & range \\ 
            \hline
            \hline        
            epochs         & [0.75d, 1.25d]\\ \hline
            learning rate  & [0.1d, d] and [d, 10d]\\ \hline
            gamma          & [0.2d, d] and [d, 3d]\\ \hline
            weight decay   & [0.2d, d] and [d, 3d]\\ \hline
            threshold      & [0, 5d]\\ \hline
        \end{tabular}
        \begin{tablenotes}
            \footnotesize
            \item The ``d'' means the default value provided by the model author. For epochs and thresholds, we use the integer value, while other hyperparameters use continuous float values.
        \end{tablenotes}
    \end{center}
\end{table}

For each mutation, we apply a mutation operator once\footnote{Due to the limitation of computing resources, we give up applying mutation operators on multiple hyperparameters.} on the original DL model structure. These new mutated models are then trained with the same dataset. We mainly focus on the mutation of two hyperparameters, i.e., epochs and learning rate, since all the models studied support these two hyperparameters. 
%used the operators on the two main hyperparameters, epochs, and learning rate. 
%Since not every model has three hyperparameters in common, 
Besides, we add a third hyperparameter for each studied model, including weight decay, gamma, and threshold. To construct mutations, we mutate three hyperparameters five times (i.e., generate five mutated DL models) for the five models studied. To overcome randomness, we train models based on each (original/mutated) DL models five times to generate five DL models. In total, we have constructed 375 (5 models$\times$3 hyperparameters$\times$5 mutations$\times$5 times) mutated models.

\subsection{Collecting metrics}
\label{sec:collection}

Our study focuses on metrics extracted from the training and evaluation of DL models, including specific energy consumption, time cost, and performance. To describe the energy consumption, we use the energy consumed by the package (pkg), main memory (ram), and the GPU (gpu), as suggested by previous research~\cite{greenAi}. We use time cost and performance to describe the training phase's performance. The cost of time refers to how much time the training phase consumes; the performance of the models is described by the accuracy metrics provided by the codes and documents. The performance metrics use the accuracy metrics offered by the model authors according to the documents. 

In this study, we employ two tools, perf~\cite{perf} and Nvidia-smi~\cite{NVIDIA-SMI} following \cite{greenAi}, to collect energy consumption and run-time data. Specifically, we list the commands in Table~\ref{tab:cmd}. 
\begin{itemize}
	\item Perf is an analyzer tool for Linux systems, using the Running Average Power Limit (RAPL~\cite{RAPL}), which is widely used~\cite{RAPL-use1, RAPL-use2, RAPL-use3} and validated by previous work~\cite{RAPL-acc4, RAPL-acc2, RAPL-acc3, RAPL-acc1}. We measure the energy consumption of CPU and RAM with run-time performance by Perf. 
	    Specifically, we use the perf with the command: 
\[{\color{blue}{\texttt{
perf stat -e power/energy-pkg/,power/energy-ram/}}}\]
\noindent which starts the training and collects the metrics of the package, RAm, and time cost. {\color{blue}\texttt{{perf stat}}} run a command and collect performance counter statistics. The {\color{blue}\texttt{{-e}}} in the command is an event selector, which is used to choose what metrics we want to collect. The energy consumption of the package, RAM is listed in this event selector.
	
	\item Nvidia-smi is a cross-platform tool that supports all standard NVIDIA driver-supported Linux distributions, providing monitoring and management functions for GPU devices. In our work, we measure the GPU energy consumption with this tool. We run the code of the training phase and collect GPU energy cost from nvidia-smi once per second. For Nvidia-smi, we run the command:
\[{\color{blue}{\texttt{
nvidia-smi --loop-ms=1000 --format=csv,noheader}}}\]
\[{\color{blue}{\texttt{--query-gpu=power.draw}}}\]
which writes the power draw of the GPU into a CSV file every second. The option, {\color{blue}{\texttt{--format=csv,noheader}}} reports the result into a CSV file without a header. {\color{blue}\texttt{{--loop-ms=1000}}} means writing metrics in the file at specified millisecond intervals. {\color{blue}{\texttt{--query-gpu=power.draw}}} tells us what information about GPU we want to report.
	
\end{itemize}

\begin{table}[t]
    \begin{center}
        \tabcolsep = 1em
        \caption{Details of the commands}
        \label{tab:cmd}
        \vspace{1ex} 
        \footnotesize
        \begin{tabular}{|c|c|c|}
            \hline
            command & parameters & metrics\\
            \hline
            \hline
            perf stat & -e power/energy-pkg/,power/energy-ram/ & pkg, ram, time\\ 
            \hline
            \multirow{3}*{nvidia-smi} & -loop-ms=1000 & \multirow{3}*{gpu} \\
            \cline{2-2}
            ~ & --format=csv,noheader & ~ \\
            \cline{2-2}
            ~ & --query-gpu=power.draw & ~ \\
            \hline
        \end{tabular}
        \begin{tablenotes}
            \footnotesize

            \item The perf command states package energy consumption and RAM energy consumption. It will report on the two former metrics with total time cost. The Nvidia command collects GPU power draw metrics every second and reports them in CSV form.
        \end{tablenotes}
    \end{center}
\end{table}

\subsection{Analyzing}
\label{sec:analysis}
The analysis methods conducted are organized in three RQs: %TODO add references for analysis methods

\textbf{RQ1:} What is the correlation between model energy consumption and their hyperparameters?

For RQ1, we employ a Spearman correlation analysis~\cite{Comparison-of-values}. We conducted a Spearman correlation analysis for each model we selected between each specific mutated hyperparameter and metrics. Then, we summarized the correlation analysis of all models and presented the results.

\textbf{RQ2:} Can we make DL models greener by adjusting hyperparameters? 

For RQ2, we argue about the trade-off between energy consumption and performance. First, we compare the metrics of each mutated model with its original. Wilcoxon signed-rank test~\cite{Connecting-historical-changes} and Cliff's delta~\cite{Cliffs-Delta-Calculator} are used to judge whether a metric from mutated models is \textit{significantly} greater or less than that of the original one. After knowing the relation between energy consumption and performance, we aim to find whether greener mutations exist, which means they decline energy consumption without performance damage by adjusting hyperparameters.

\textbf{RQ3: Are the conclusions above able to apply to the parallel environment?}

For RQ3, we investigate the difference when multiple models are trained in a parallel environment. Based on our computing resources, we train two models in parallel once. First, we run a model in the background; then, we mutate and train a model to collect data. The data are analyzed in the same way as RQ1 and RQ2. To show the difference, we draw heat maps with the distinction between metrics in parallel and not in parallel.

\section{Experimental results}
\label{sec:result}
In this section, we specify the results of our RQs, along with our motivations, details of approaches, and findings.

\begin{table}[t]
    \begin{center}
        \tabcolsep = 1em
        \centering
        \caption{Spearman correlation single energy}
        \vspace{1ex}
        \label{tab:1}
        \footnotesize
        \begin{tabular}{*{5}{|c}|}
            \hline
            hyperparameter & package & ram   & gpu   & total\\ \hline   
            \hline
            epochs:  & 0/0/0/0/6 & 0/0/0/0/6 & 0/0/0/0/6 & 0/0/0/0/18\\ \hline
            learning rate:  & 0/0/6/0/0 & 0/1/4/1/0 & 0/3/3/0/0 & 0/4/13/1/0\\ \hline
            weight decay:  & 0/0/3/0/0 & 0/1/1/1/0 & 0/1/2/0/0 & 0/2/6/1/0\\ \hline
            gamma:  & 0/0/2/0/0 & 0/0/2/0/0 & 0/0/2/0/0 & 0/0/6/0/0\\ \hline
            threshold:  & 0/0/1/0/0 & 0/0/1/0/0 & 0/0/0/1/0 & 0/0/2/1/0\\ \hline
            total: & 0/0/12/0/6 & 0/2/8/2/6 & 0/4/7/1/6 & 0/6/27/3/18\\ \hline
        \end{tabular}
        \begin{tablenotes}
            \footnotesize
            \item Spearman
        \end{tablenotes}
    \end{center}
\end{table}

\begin{table}[t]
    \begin{center}
        \tabcolsep = 1em
        \centering
        \caption{Spearman correlation single performance}
        \vspace{1ex}
        \label{tab:1}
        \footnotesize
        \begin{tabular}{*{4}{|c}|}
            \hline
            hyperparameter & time  & performance & total\\ \hline   
            \hline
            epochs:  & 0/0/0/0/6 & 0/0/4/0/1 & 0/0/4/0/7\\ \hline
            learning rate:  & 0/1/5/0/0 & 0/4/1/1/0 & 0/5/6/1/0\\ \hline
            weight decay:  & 0/0/3/0/0 & 0/0/2/0/0 & 0/0/5/0/0\\ \hline
            gamma:  & 0/0/2/0/0 & 0/0/2/0/0 & 0/0/4/0/0\\ \hline
            threshold:  & 0/0/1/0/0 & 0/0/1/0/0 & 0/0/2/0/0\\ \hline
            total: & 0/1/11/0/6 & 0/4/10/1/1 & 0/5/21/1/7\\ \hline
        \end{tabular}
        \begin{tablenotes}
            \footnotesize
            \item Spearman
        \end{tablenotes}
    \end{center}
\end{table}

\begin{table}[t]
    \begin{center}
        \tabcolsep = 1em
        \centering
        \caption{Spearman correlation parallel energy}
        \vspace{1ex}
        \label{tab:1}
        \footnotesize
        \begin{tabular}{*{6}{|c}|}
            \hline
            hyperparameter & package & ram   & gpu   & total\\ \hline   
            \hline
            epochs:  & 0/0/0/0/10 & 0/0/0/0/10 & 0/0/0/0/10 & 0/0/0/0/30\\ \hline
            learning rate:  & 0/0/10/0/0 & 0/0/9/1/0 & 1/5/4/0/0 & 1/5/23/1/0\\ \hline
            weight decay:  & 0/1/5/0/0 & 0/1/4/1/0 & 0/1/4/1/0 & 0/3/13/2/0\\ \hline
            gamma:  & 0/1/3/0/0 & 0/1/3/0/0 & 0/1/3/0/0 & 0/3/9/0/0\\ \hline
            total: & 0/2/18/0/10 & 0/2/16/2/10 & 1/7/11/1/10 & 1/11/45/3/30\\ \hline
        \end{tabular}
        \begin{tablenotes}
            \footnotesize
            \item Spearman
        \end{tablenotes}
    \end{center}
\end{table}

\begin{table}[t]
    \begin{center}
        \tabcolsep = 1em
        \centering
        \caption{Spearman correlation single performance}
        \vspace{1ex}
        \label{tab:1}
        \footnotesize
        \begin{tabular}{*{4}{|c}|}
            \hline
            hyperparameter & time  & performance & total\\ \hline   
            \hline
            epochs:  & 0/0/0/0/10 & 0/0/8/1/0 & 0/0/8/1/10\\ \hline
            learning rate:  & 0/1/9/0/0 & 0/5/4/1/0 & 0/6/13/1/0\\ \hline
            weight decay:  & 0/1/5/0/0 & 0/1/3/0/0 & 0/2/8/0/0\\ \hline
            gamma:  & 0/1/3/0/0 & 0/0/3/1/0 & 0/1/6/1/0\\ \hline
            total: & 0/3/17/0/10 & 0/6/18/3/0 & 0/9/35/3/10\\ \hline
        \end{tabular}
        \begin{tablenotes}
            \footnotesize
            \item Spearman
        \end{tablenotes}
    \end{center}
\end{table}

\begin{table}[t]
    \begin{center}
        \tabcolsep = 1em
        \centering
        \caption{Trade-off single}
        \vspace{1ex}
        \label{tab:1}
        \footnotesize
        \begin{tabular}{*{6}{|c}|}
            \hline
            hyperparameter & model amount & package & ram & gpu & total trade-off\\ \hline
            \hline
            epochs         & 6 & 6 & 5 & 5 & 16\\ \hline
            learning rate  & 6 & 4 & 4 & 3 & 11\\ \hline
            weight decay   & 3 & 0 & 2 & 0 & 2\\ \hline
            gamma          & 2 & 1 & 0 & 2 & 3\\ \hline
            threshold      & 1 & 0 & 1 & 2 & 1\\ \hline
            total          & - & 12 & 11 & 11 & 34\\ \hline
        \end{tabular}
        \begin{tablenotes}
            \footnotesize
            \item Causal
        \end{tablenotes}
    \end{center}
\end{table}

\begin{table}[t]
    \begin{center}
        \tabcolsep = 1em
        \centering
        \caption{Trade-off parallel}
        \vspace{1ex}
        \label{tab:1}
        \footnotesize
        \begin{tabular}{*{6}{|c}|}
            \hline
            hyperparameter & model amount & package & ram & gpu & total trade-off \\ \hline
            \hline
            epochs         & 10 & 8 & 6 & 6 & 20\\ \hline
            learning rate  & 10 & 5 & 5 & 9 & 19\\ \hline
            weight decay   & 6 & 4 & 3 & 3 & 10\\ \hline
            gamma          & 4 & 1 & 1 & 1 & 3 \\ \hline
            total          & - & 18 & 15 & 19 & 52\\ \hline
        \end{tabular}
        \begin{tablenotes}
            \footnotesize
            \item Causal
        \end{tablenotes}
    \end{center}
\end{table}

\begin{table}[t]
    \begin{center}
        \tabcolsep = 1em
        \centering
        \caption{influence single}
        \vspace{1ex}
        \label{tab:1}
        \footnotesize
        \begin{tabular}{*{6}{|c}|}
            \hline
            model & epochs & learning rate & weight decay & gamma & threshold\\ \hline
            \hline
            mst	 & 0.9281 & -0.0090 & - & 0.0079 & -\\ \hline
            mff	 & 0.9853 & -0.0378 & - & - & 0.0066\\ \hline
            sia	 & 0.9332 & -0.0357 & - & 0.0166 & -\\ \hline
            ca	 & 0.7077 & 0.4179 & -0.1892 & - & -\\ \hline
            de	 & 0.9756 & -0.1247 & -0.0099 & - & -\\ \hline
            hr	 & 0.9718 & 0.0159 & -0.0181 & - & -\\ \hline
        \end{tabular}
        \begin{tablenotes}
            \footnotesize
            \item OLS R2: ca 0.53
        \end{tablenotes}
    \end{center}
\end{table}

\begin{table}[t]
    \begin{center}
        \tabcolsep = 1em
        \centering
        \caption{influence parallel}
        \vspace{1ex}
        \label{tab:1}
        \footnotesize
        \begin{tabular}{*{5}{|c}|}
            \hline
            model & epochs & learning rate & weight decay & gamma \\ \hline
            \hline
            mst\_sia	 & 0.7328 & -0.0023 & - & -0.0002 \\ \hline
            mst\_ca	 & 0.8043 & -0.0652 & - & 0.0641 \\ \hline
            sia\_mst	 & 0.9617 & -0.0460 & - & -0.0315\\ \hline
            sia\_ca	 & 0.8529 & -0.0424 & - & -0.0507 \\ \hline
            ca\_mst	 & 0.9572 & -0.0573 & -0.0235 & - \\ \hline
            ca\_sia	 & 0.9407 & -0.0725 & 0.0401 & - \\ \hline
            ca\_de	 & 0.8732 & -0.0378 & -0.0029 & -\\ \hline
            de\_ca	 & 0.9036 & -0.0907 & 0.0240 & - \\ \hline
            de\_hr	 & 0.4428 & -0.0095 & -0.0169 & -\\ \hline
            hr\_de	 & 0.8168 & -0.1276 & 0.1404 & -\\ \hline
        \end{tabular}
        \begin{tablenotes}
            \footnotesize
            \item OLS R2: sia\_ca 0.84, hr\_de 0.84, mst\_sia 0.71, de\_hr 0.62, mst\_ca 0.59
        \end{tablenotes}
    \end{center}
\end{table}

\subsection{RQ1: correlation between energy consumption and hyperparameters}
\noindent \textbf{Motivation and Approach.}
Our goal is to investigate the impact of hyperparameters on the performance and energy consumption of a DL model during its training phase. %In addition to adjusting the hyperparameters to check the accuracy of the model, we must also consider the potential changes in energy consumption resulting from these modifications. 
In this RQ, we performed a Spearman correlation analysis between common hyperparameters and the metrics extracted from the training phase for five chosen DL models and their reputations. The hyperparameters we chose consist of epochs, learning rate, gamma, weight decay, and threshold. %Since different models apply different hyperparameters, we choose the most common ones from all five models. 

\noindent \textbf{Result.}
The result of the Spearman correlation analysis on five models is summarized in Table~\ref{tab:1}. The numbers in the table present whether the hyperparameters have a significant correlation with specific metrics. The first number in each grid means the number of positive correlations (the Spearman coefficient $ \rho \ge 0.2$ and $p < 0.05$) for the five models; the second number shows the number of no significant correlations ($ -0.2\le \rho \le 0.2$ or $p \ge 0.05$); the third number conveys the number of negative correlations ($ \rho \le -0.2$ and $p < 0.05$). 
% (Positive correlation: $ \rho \ge 0.2$ and $p < 0.05$; No significant correlation: $ \rho \le 0.2$ or $ \rho \ge -0.2$ or $p \ge 0.05$; Negative correlation: $ \rho \le -0.2$ and $p < 0.05$). 

\begin{table}[t]
    \begin{center}
        \tabcolsep = 1em
        \centering
        \caption{The summary of Spearman correlation analysis on the training phase of five hyperparameters.}
        \vspace{1ex}
        \label{tab:1}
        \footnotesize
        \begin{tabular}{*{6}{|c}|}
            \hline
            hyperparameter & package & ram   & gpu   & time  & performance\\ \hline   
            \hline
            epochs         & 5/0/0   & 5/0/0 & 4/1/0 & 5/0/0 & 1/4/0\\ \hline
            learning rate  & 0/5/0   & 0/4/1 & 0/2/3 & 0/4/1 & 2/1/2\\ \hline
            gamma          & 0/1/1   & 0/1/1 & 0/1/1 & 0/2/0 & 1/1/0\\ \hline
            weight decay   & 0/2/0   & 1/0/1 & 0/2/0 & 0/2/0 & 0/2/0\\ \hline
            threshold      & 0/1/0   & 0/1/0 & 1/0/0 & 0/1/0 & 0/1/0\\ \hline
        \end{tabular}
        \begin{tablenotes}
            \footnotesize
            \item The values in the table show the sum of five correlation coefficients. The Spearman correlation coefficients are between hyperparameters and metrics on energy consumption,  time, and performance of the models. A positive correlation means the Spearman correlation coefficient $\ge 0.2$ and $p$-value $< 0.05$. A negative correlation means the Spearman correlation coefficients $\le -0.2$ and the $p$-value $< 0.05$. The other situation has no significant correlation. For example, ``1/4/0'' stands for 1 positive correlation, 4 no significant correlations, and no negative correlation.
        \end{tablenotes}
    \end{center}
\end{table}

Based on the data presented in Table~\ref{tab:1}, we can see that there is a strong positive correlation between the number of epochs and all energy consumption metrics, as well as time cost. However, there is only a weak correlation between the number of epochs and performance. This suggests that by adjusting the number of epochs appropriately, we can reduce energy consumption without compromising accuracy.

As for the learning rate, there are weaker correlations with energy consumption and time cost but a higher correlation with training performance than the epochs. We can also observe that mutations in learning rate have a negative relationship with RAM, GPU, and run-time, especially with GPU energy consumption. Hence, it is more likely to harm the training performance if we plan to decrease energy consumption by mutating the learning rate.

The third studied hyperparameters are various for the five studied models. As seen in Table~\ref{tab:1}, Gamma negatively correlates with three energy metrics. The others are unrelated to most metrics, but weight decay is weakly related to RAM. Therefore, modifying these hyperparameters declines specific energy consumption.

\begin{framed}
        \noindent\textbf{Answer to RQ1.} Many hyperparameters have a positive/negative correlation with energy consumption. Specifically, epochs have a significant positive correlation with both energy consumption and time cost. On the other hand, the learning rate has a weak negative correlation with RAM, GPU, and time cost. 
\end{framed}

\subsection{RQ2: greener DL models by adjusting hyperparameters}
\noindent \textbf{Motivation and Approach.}
%consumption for model performance by adjusting certain hyperparameters. 
Once we establish the correlation, as shown in RQ 1, we aim to find a better balance while modifying the hyperparameters.
%A model that consumes less energy should deliver better performance or consume less energy while delivering similar performance. Therefore, 
Specifically, our goal is to obtain the mutations leading to a mutated model that is more energy efficient than the original model (i.e., performs better with similar energy or consumes less energy with similar performance).

To describe the differences between mutations and origins, we employ the Wilcoxon signed-rank test and Cliff's delta to judge whether our energy consumption and performance metrics are significantly greater or less than their original values. Three relationships are defined, namely ``win'', ``tie'', and ``loss''. The ``win'' means the metrics are significantly greater, i.e., $p < 0.05$ and $\delta \ge 0.147$. The ``tie'' means that the metrics are not significantly greater or less, i.e., $p \ge 0.05$ or $-0.147 < \delta < 0.147$. The ``loss'' means that the metrics are significantly less, i.e., $p < 0.05$ and $\delta \le -0.147$).

\begin{figure}[t]
    \centering
    \begin{minipage}{0.32\textwidth}
        \centering
        \includegraphics[scale=0.17]{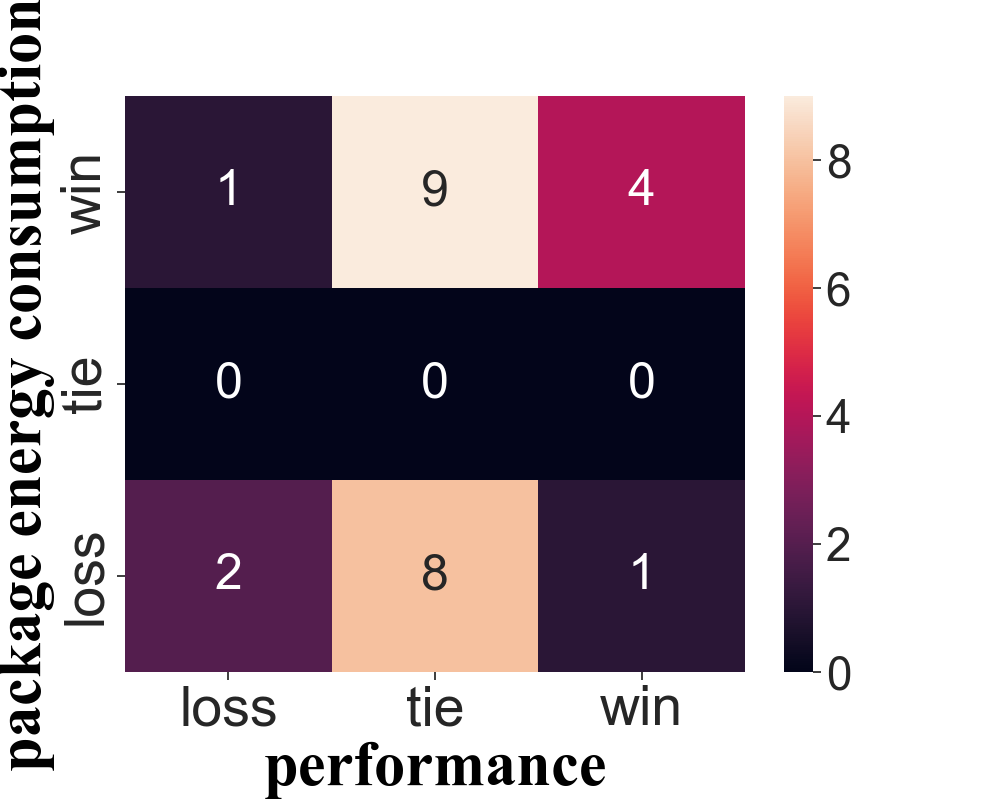}
        (a) package energy consumption
    \end{minipage}
    \begin{minipage}{0.32\textwidth}
        \centering
        \includegraphics[scale=0.17]{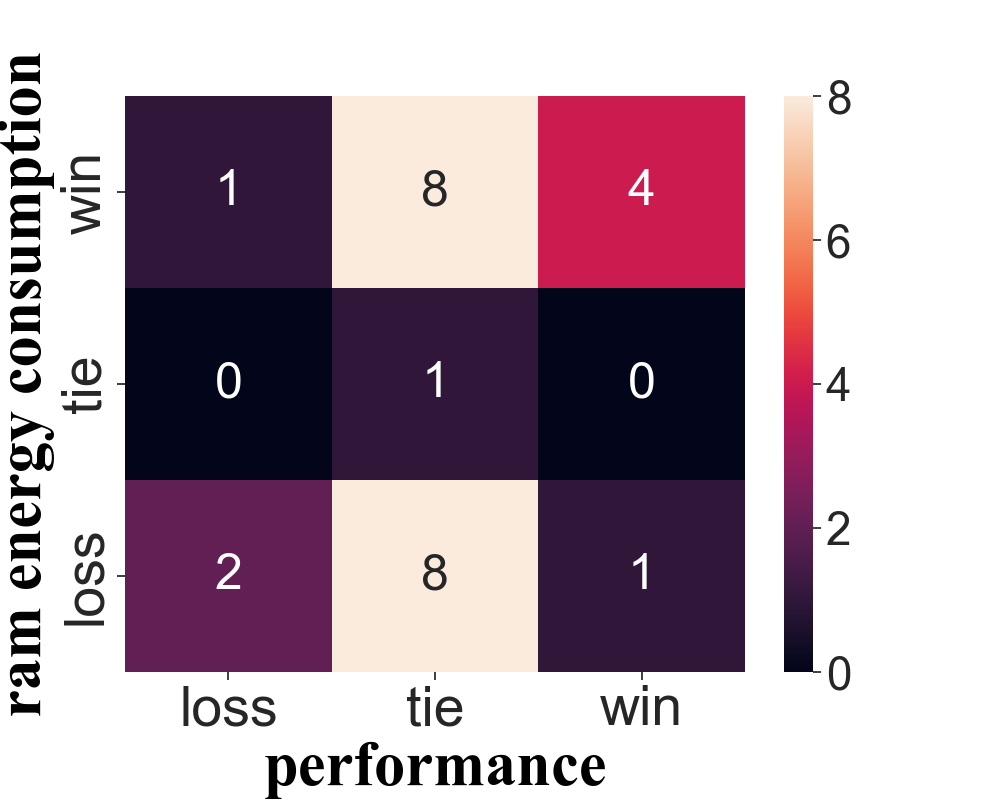}
        (b) ram energy consumption
    \end{minipage}
    \begin{minipage}{0.32\textwidth}
        \centering
        \includegraphics[scale=0.17]{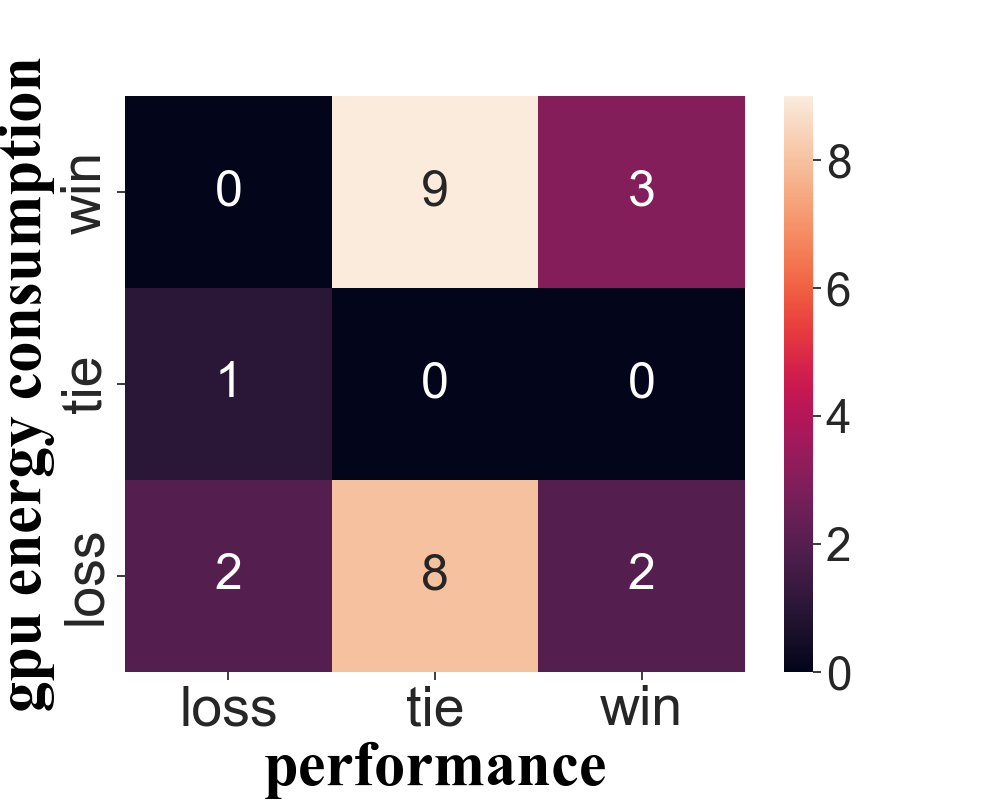}
        (c) gpu energy consumption
    \end{minipage}
    \caption{Trade-off between energy consumption and performance when mutating epochs. Numbers in each grid mean how many mutated models ``win-tie-loss'' in energy consumption and network performance. For example, 1 in Figure (a) shows only one model ``wins'' in package energy consumption and ``loses'' in network performance.}
    \label{fig:epoch}
\end{figure}

\begin{figure}[t]
    \centering
    \begin{minipage}{0.32\textwidth}
        \centering
        \includegraphics[scale=0.17]{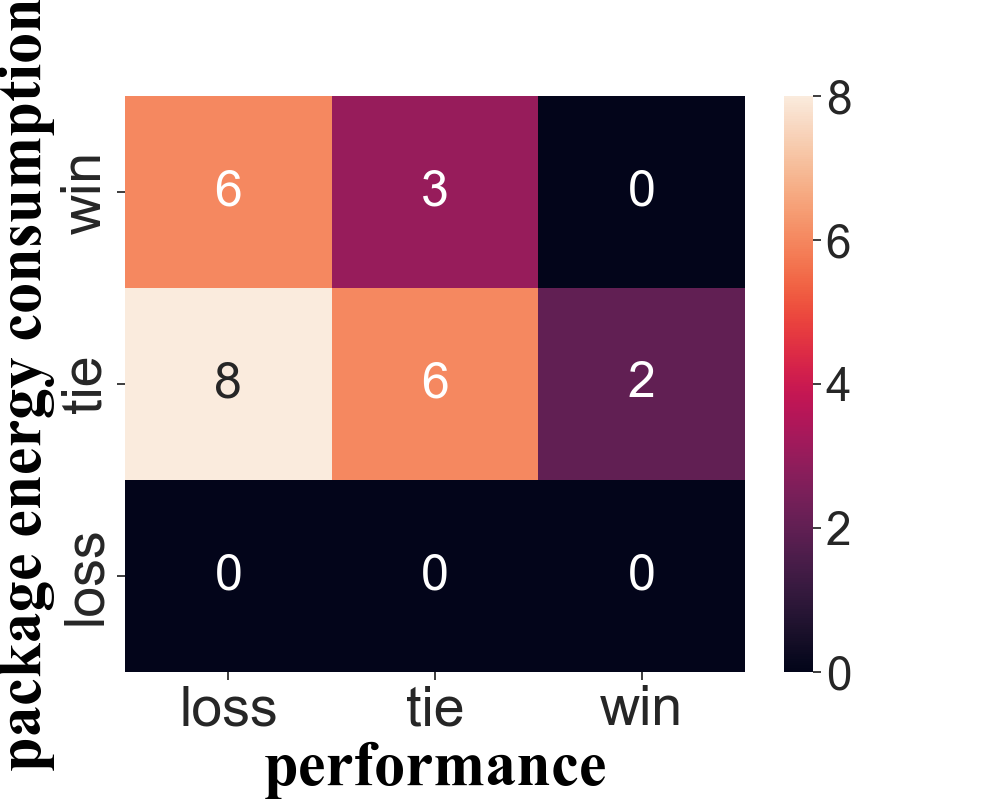}
        (a) package energy consumption
    \end{minipage}
    \begin{minipage}{0.32\textwidth}
        \centering
        \includegraphics[scale=0.17]{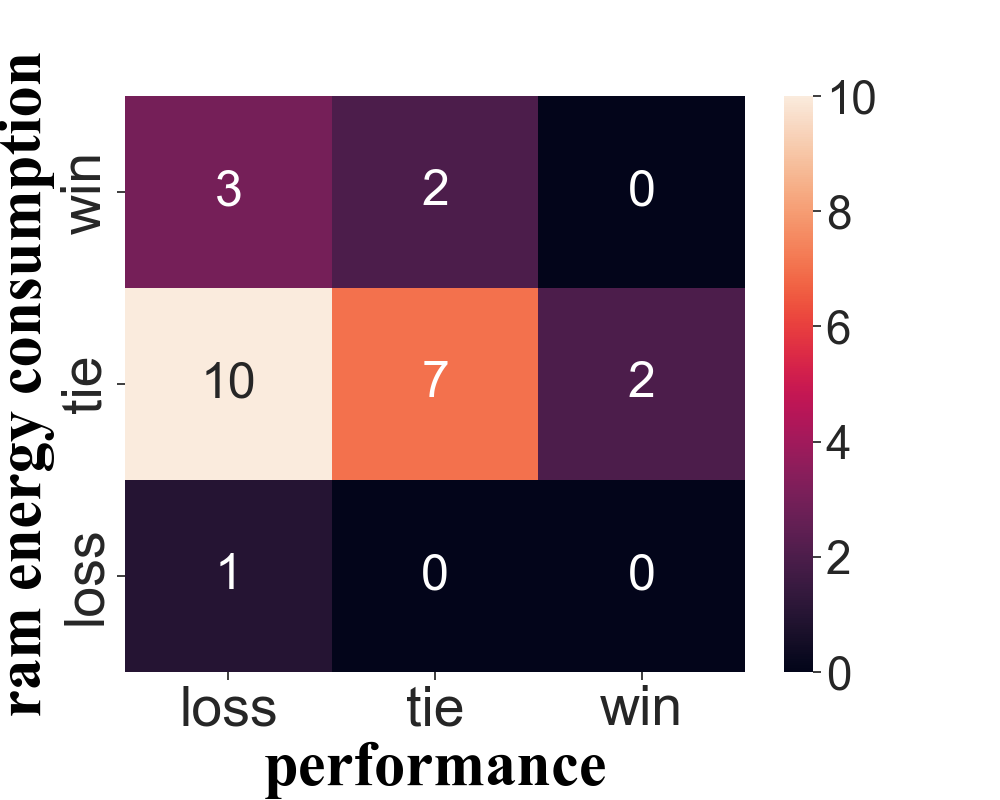}
        (b) ram energy consumption
    \end{minipage}
    \begin{minipage}{0.32\textwidth}
        \centering
        \includegraphics[scale=0.17]{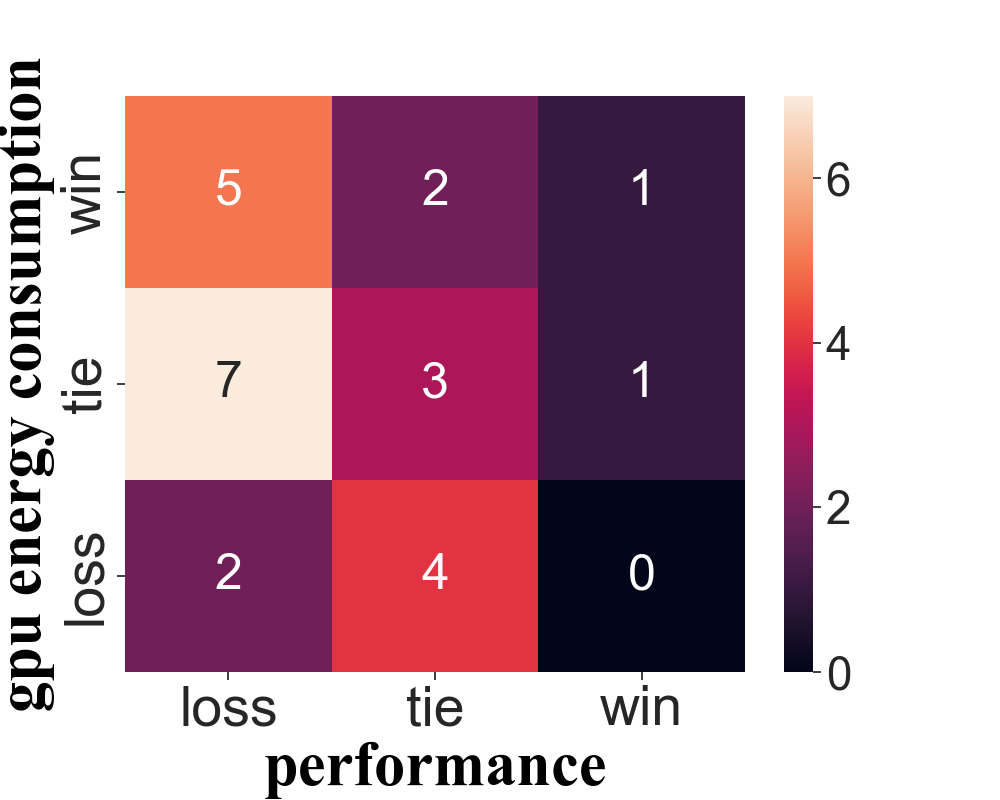}
        (c) gpu energy consumption
    \end{minipage}
    \caption{Trade off between energy consumption and performance when mutating learning rate}
    \label{fig:lr}
\end{figure}

\begin{figure}[t]
    \centering
    \begin{minipage}{0.32\textwidth}
        \centering
        \includegraphics[scale=0.17]{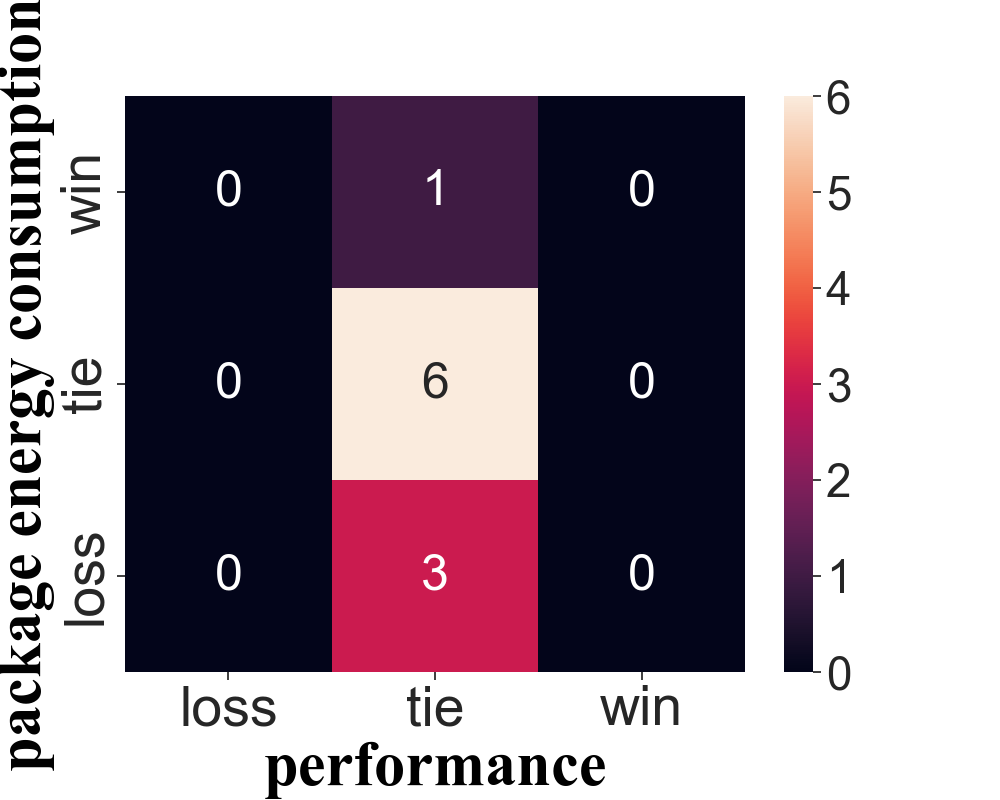}
        (a) package energy consumption
    \end{minipage}
    \begin{minipage}{0.32\textwidth}
        \centering
        \includegraphics[scale=0.17]{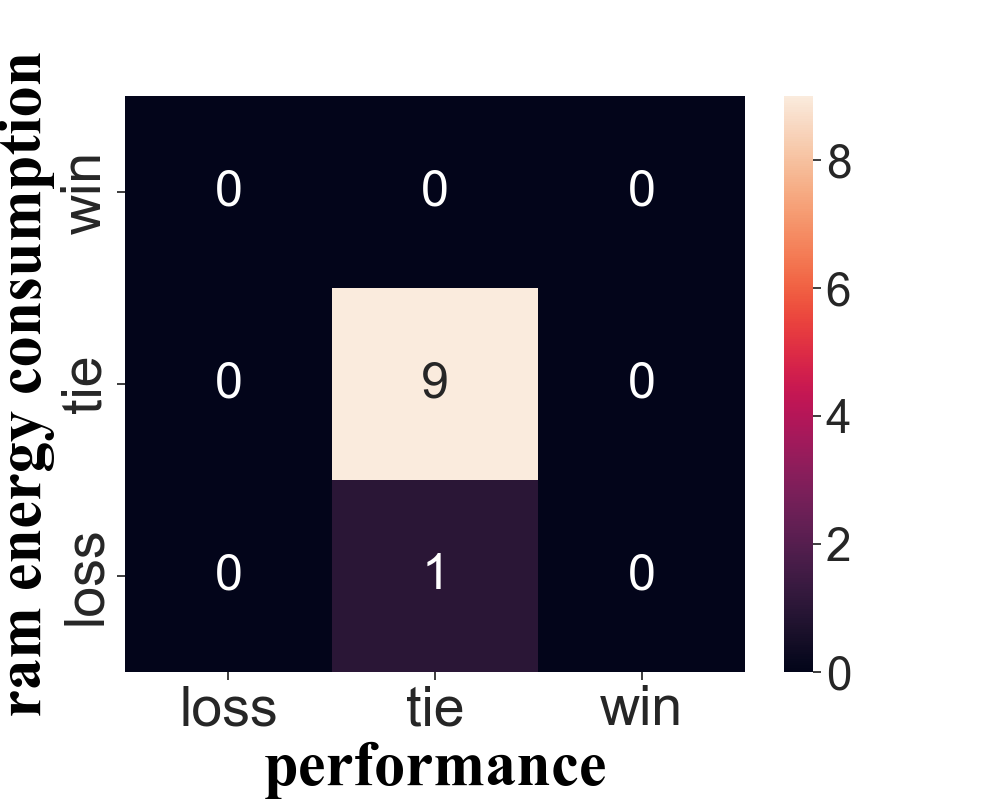}
        (b) ram energy consumption
    \end{minipage}
    \begin{minipage}{0.32\textwidth}
        \centering
        \includegraphics[scale=0.17]{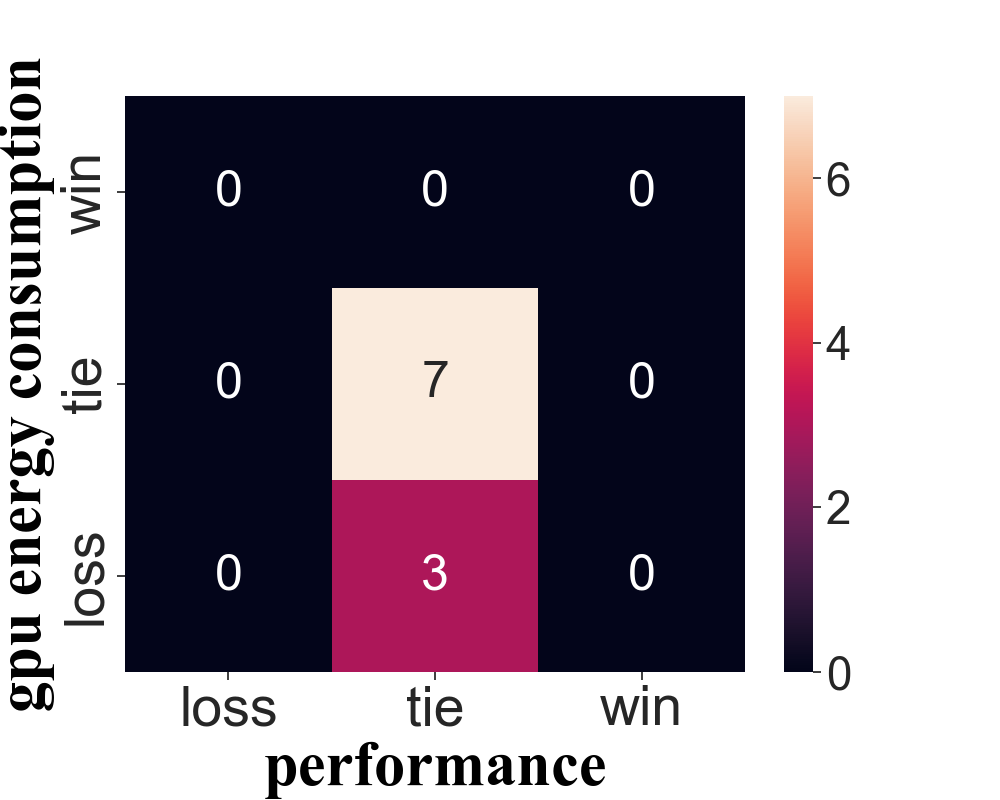}
        (c) gpu energy consumption
    \end{minipage}
    \caption{Trade off between energy consumption and performance when mutating gamma}
    \label{fig:gamma}
\end{figure}

\begin{figure}[t]
    \centering
    \begin{minipage}{0.32\textwidth}
        \centering
        \includegraphics[scale=0.17]{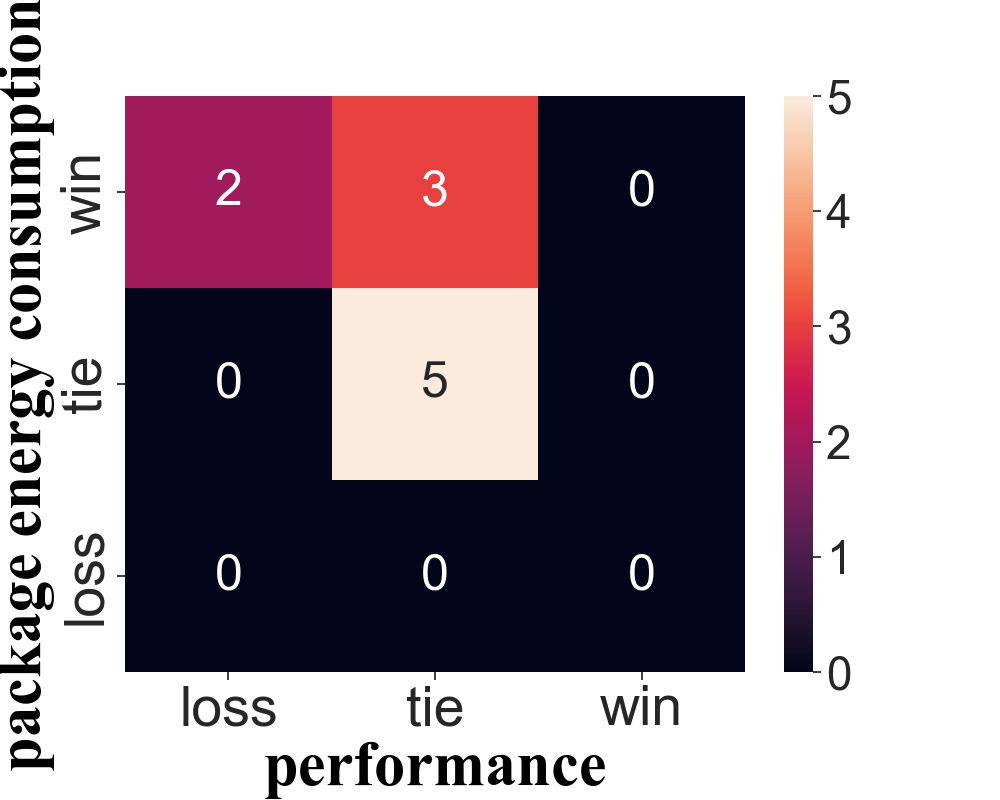}
        (a) package energy consumption
    \end{minipage}
    \begin{minipage}{0.32\textwidth}
        \centering
        \includegraphics[scale=0.17]{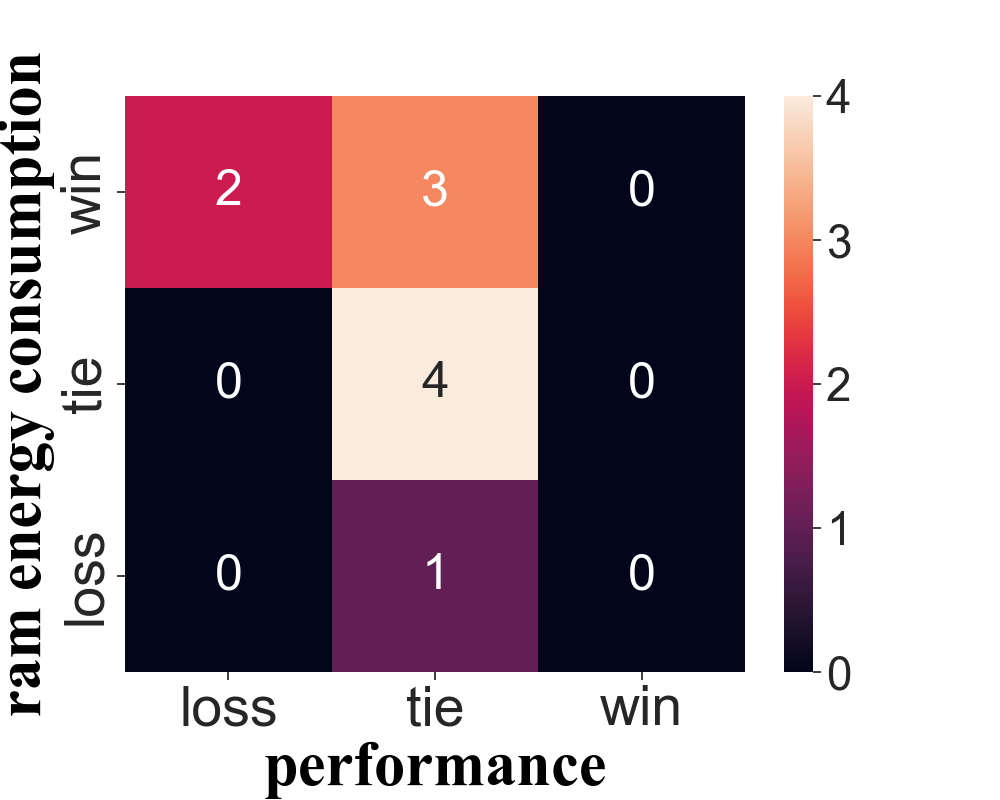}
        (b) ram energy consumption
    \end{minipage}
    \begin{minipage}{0.32\textwidth}
        \centering
        \includegraphics[scale=0.17]{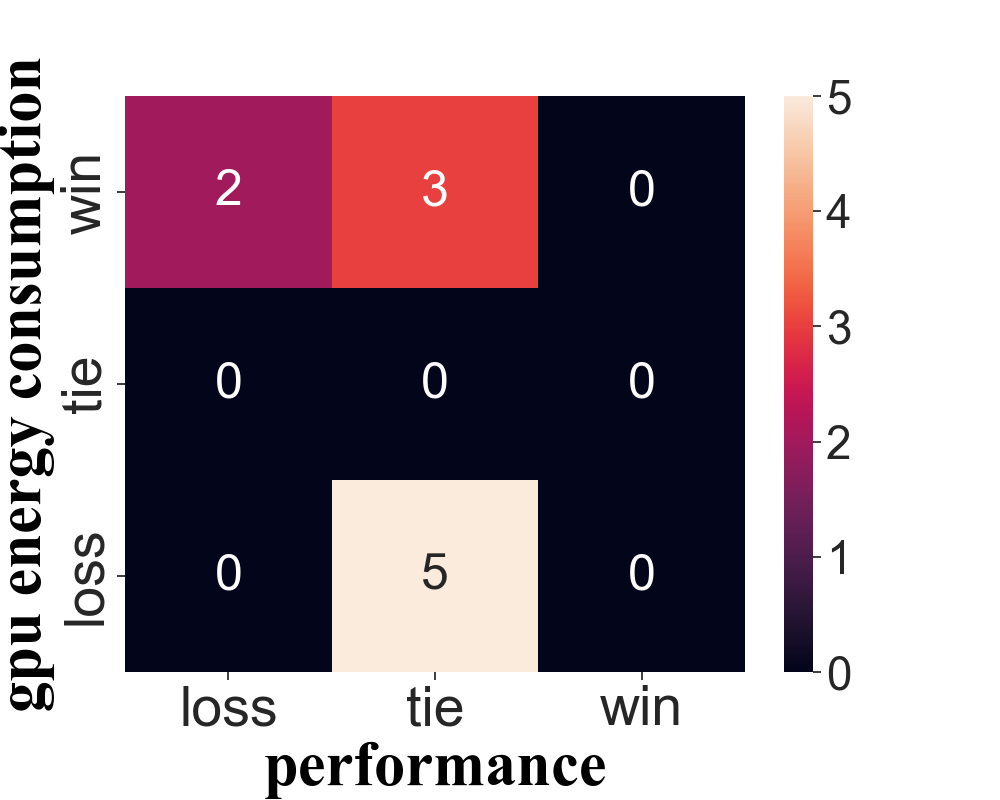}
        (c) gpu energy consumption
    \end{minipage}
    \caption{Trade off between energy consumption and performance when mutating weight decay}
    \label{fig:wd}
\end{figure}

\begin{figure}[t]
    \centering
    \begin{minipage}{0.32\textwidth}
        \centering
        \includegraphics[scale=0.17]{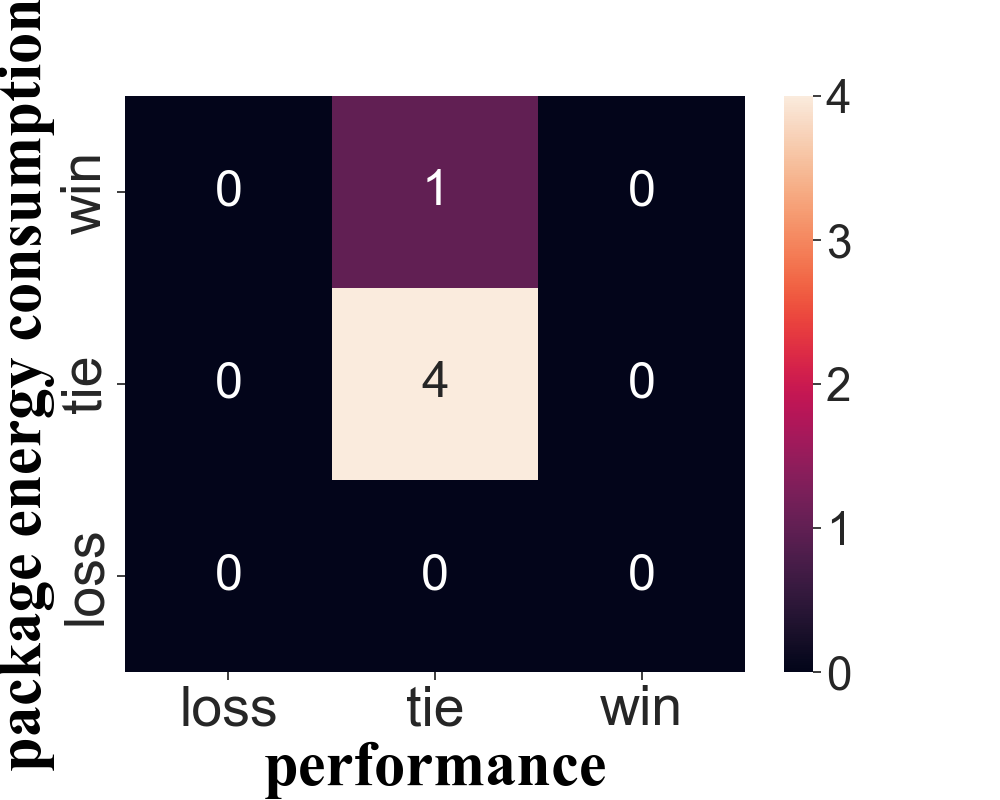}
        (a) package energy consumption
    \end{minipage}
    \begin{minipage}{0.32\textwidth}
        \centering
        \includegraphics[scale=0.17]{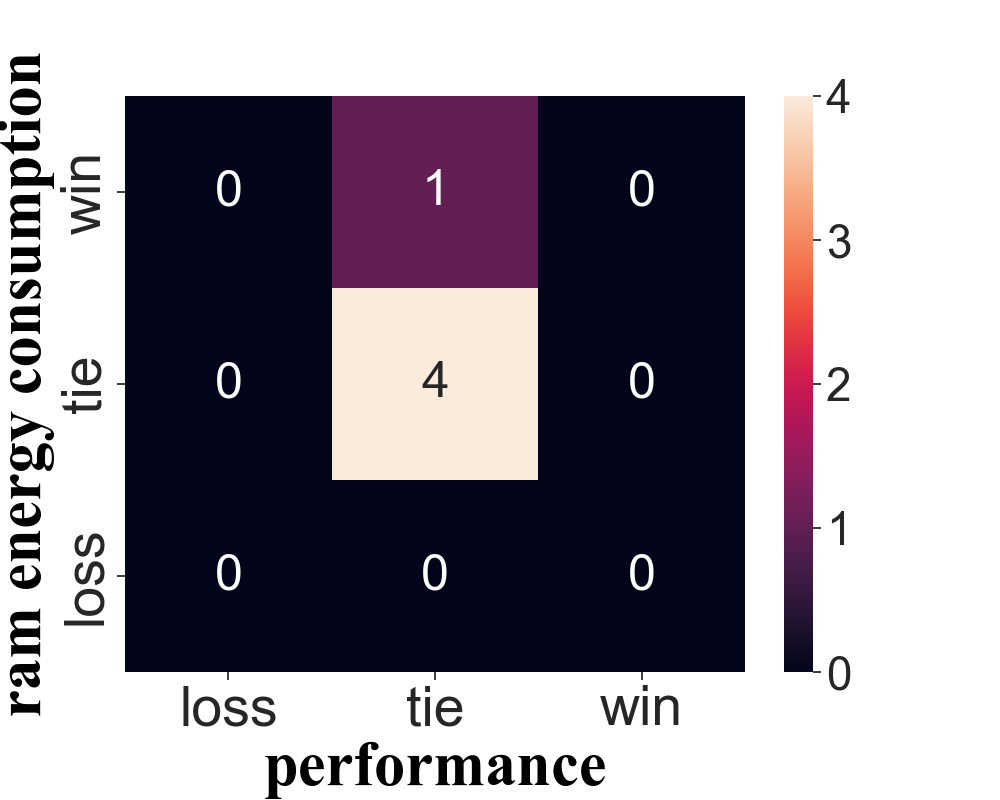}
        (b) ram energy consumption
    \end{minipage}
    \begin{minipage}{0.32\textwidth}
        \centering
        \includegraphics[scale=0.17]{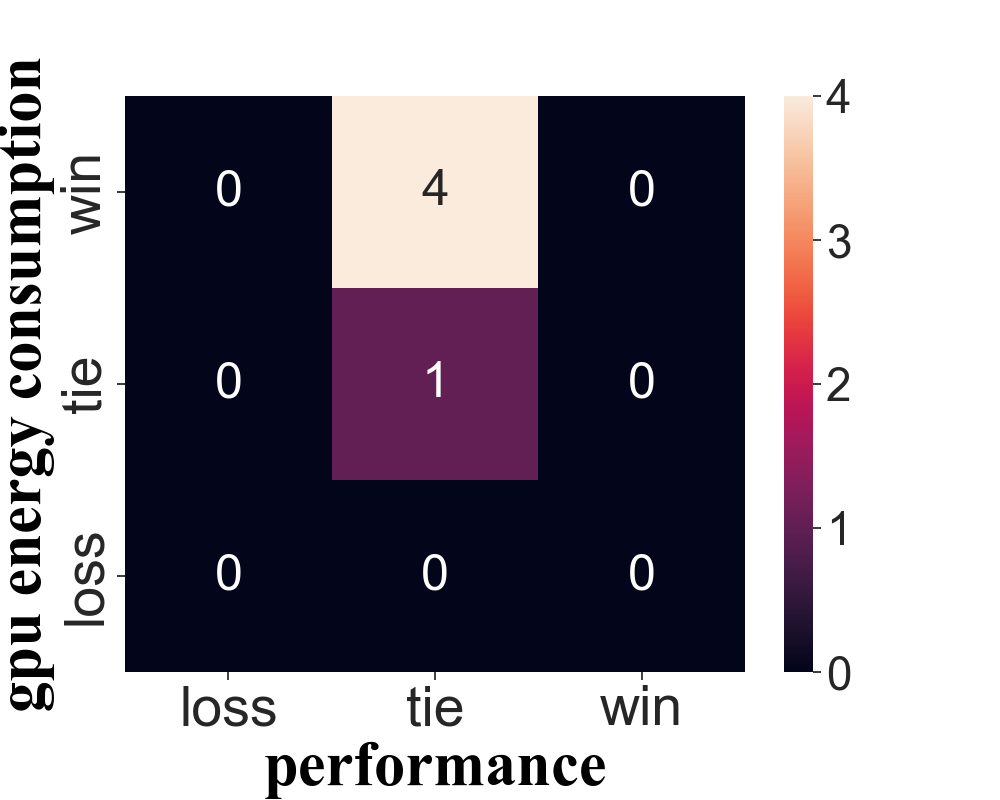}
        (c) gpu energy consumption
    \end{minipage}
    \caption{Trade off between energy consumption and performance when mutating threshold}
    \label{fig:t}
\end{figure}

\noindent \textbf{Result.}
The results of the trade-off analysis are organized in the following five figures (i.e., Figures~\ref{fig:epoch}-\ref{fig:t}) for five DL hyperparameters. Each figure consists of three heat maps for three energy metrics. 
%Specifically, Figure~\ref{fig:epoch} presents the trade-off between the epochs and energy metrics. Figure~\ref{fig:lr} shows the analysis results when mutating the learning rate. Figure~\ref{fig:gamma} shows the trade-off between the gamma and energy consumption. Figure~\ref{fig:wd} shows the result of the weight-decay trade-off analysis. Figure~\ref{fig:t} presents the result between the threshold and the energy consumption metrics.
In the heat maps, each grid represents a special type of ``win-tie-loss''. The $y$-axis shows whether the energy consumption of mutations is greater or lower than that of the original models. The $x$-axis reveals whether the performance is better or worse than the original. For instance, Figure~\ref{fig:epoch}(a) means there is one mutation (i.e., mutated model structure) that ``wins'' in package energy consumption and ``loses'' in performance; nine mutations ``win'' in energy cost and ``tie'' in performance; four ``wins'' in both energy consumption and performance; two costs less energy and gains worse performance; eight uses less energy than default epochs and obtains similar performance; and one costs less energy and gains better performance.

The trade-off between the epochs and energy metrics is presented in Figure~\ref{fig:epoch}. The figure shows that when epochs are modified in a limited range, new training phases are more likely to draw a ``tie'' in performance but ``win'' or ``lose'' in energy consumption. Among the three types of energy metrics, there are no notable differences. In most test models, reducing the epochs in a suitable step can significantly reduce the cost of energy and does not harm the performance.

Figure~\ref{fig:lr} shows the results of the trade-off analysis when mutating the learning rate. The figure shows that mutating the learning rate randomly harms the training results most of the time, which ``wins'' in energy consumption and ``loses'' in performance. However, there are still some models that ``win'' in performance or ``lose'' in energy consumption. Therefore, adjusting with purpose can achieve a greener training phase compared to randomly mutating the learning rate. Especially for GPU energy consumption, it is more likely that ``lose'' in energy consumption while ``tie'' in performance; hence, for some models that mainly consume energy on GPU, modifying the learning rate can decrease the total energy cost.

Figure~\ref{fig:gamma} shows the result of the gamma. We can observe that all results have similar accuracy metrics; hence, changing the value of gamma has little effect on performance. Meanwhile, it frequently affects the energy costs of the network. Therefore, adjusting gamma carefully can lower the energy consumption metrics.

Figure~\ref{fig:wd} presents the weight decay trade-off analysis. It is shown that mutating weight decay has a weak impact on model performance because the performance of most mutations is similar to the original models. For energy cost, package and RAM are increased or made no difference by the mutation, while GPU energy consumption declines for half of the test samples. Adjusting weight decay can cost less if GPU mainly costs the model's energy.

Figure~\ref{fig:t} presents the trade-off result between the threshold and the energy consumption metrics. The figure shows that the mutating threshold ``wins'' in energy consumption, but it ``ties'' in performance. Especially for GPU metrics, more models cost more energy. Therefore, adjusting the threshold may not lead to a greener training phase.

\begin{framed}
        \noindent\textbf{Answer to RQ2.} Modifying hyperparameters would gain better trade-offs. Adjusting epochs would lead to less energy, and adjusting the learning rate ``wins'' in performance. For GPU energy consumption, mutating weight decay and learning rate can lead to less energy cost.
\end{framed}

\subsection{RQ3: the parallel environment}

\noindent \textbf{Motivation and Approach.}
In the above RQs, each model is trained separately.
%In this RQ, we aim to find out whether our conclusions above work in parallel environments. 
Usually, more than one model is trained in parallel, especially on shared servers.  We want to know whether our conclusions from previous RQs can be applied to parallel environments. Therefore, we run two models in pairs to imitate the parallel environment and redo the experiments of the former RQs to find differences. In our parallel environment, one model is mutated and trained in the background, and the other model is mutated and measured to collect metrics. Since some errors occur in several pairs, the data from four pairs is successfully collected and finally analyzed.

\begin{figure}[t]
    \centering
    \begin{minipage}{0.32\textwidth}
        \centering
        \includegraphics[scale=0.17]{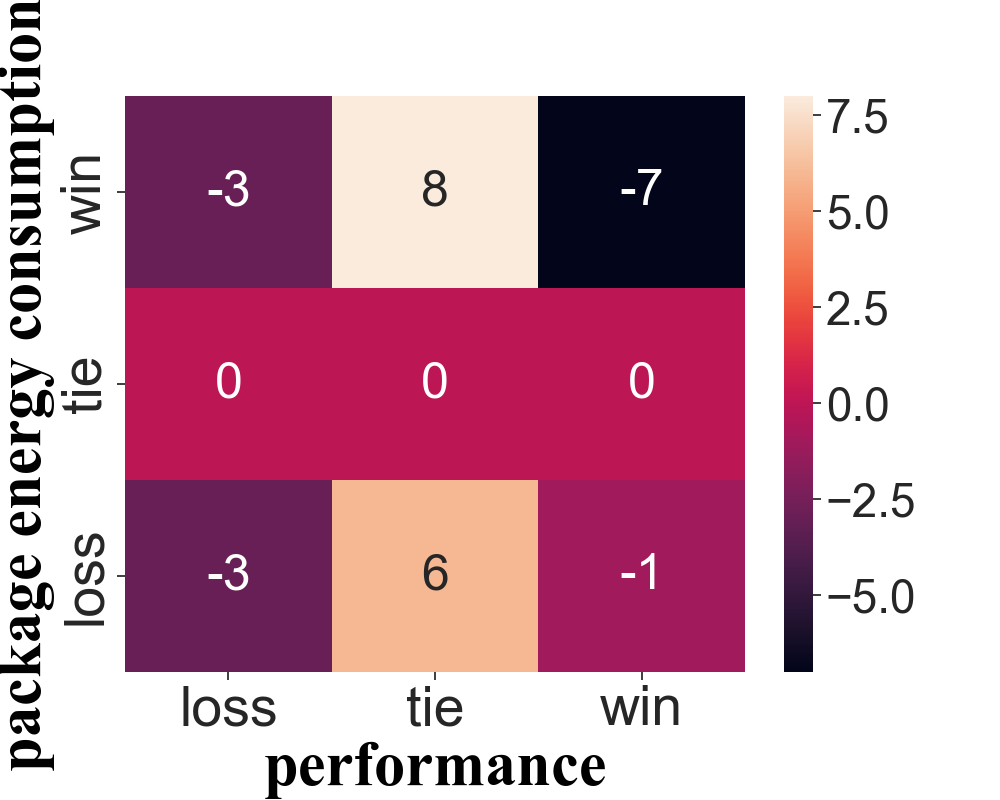}
        (a) package energy consumption
    \end{minipage}
    \begin{minipage}{0.32\textwidth}
        \centering
        \includegraphics[scale=0.17]{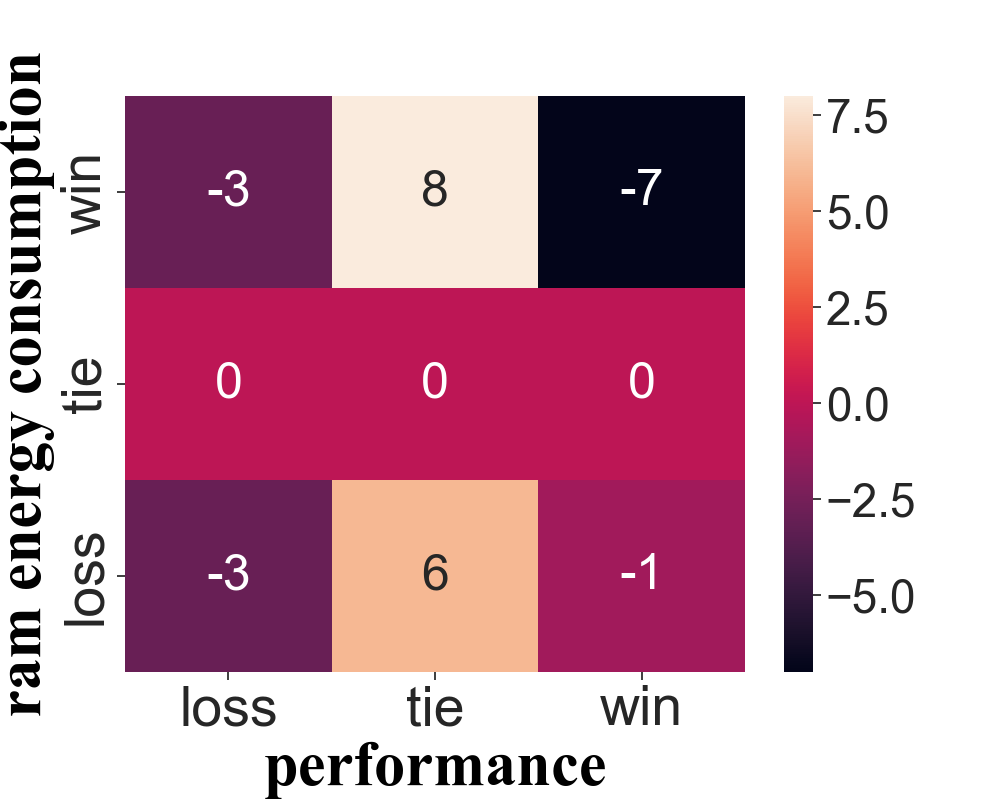}
        (b) ram energy consumption
    \end{minipage}
    \begin{minipage}{0.32\textwidth}
        \centering
        \includegraphics[scale=0.17]{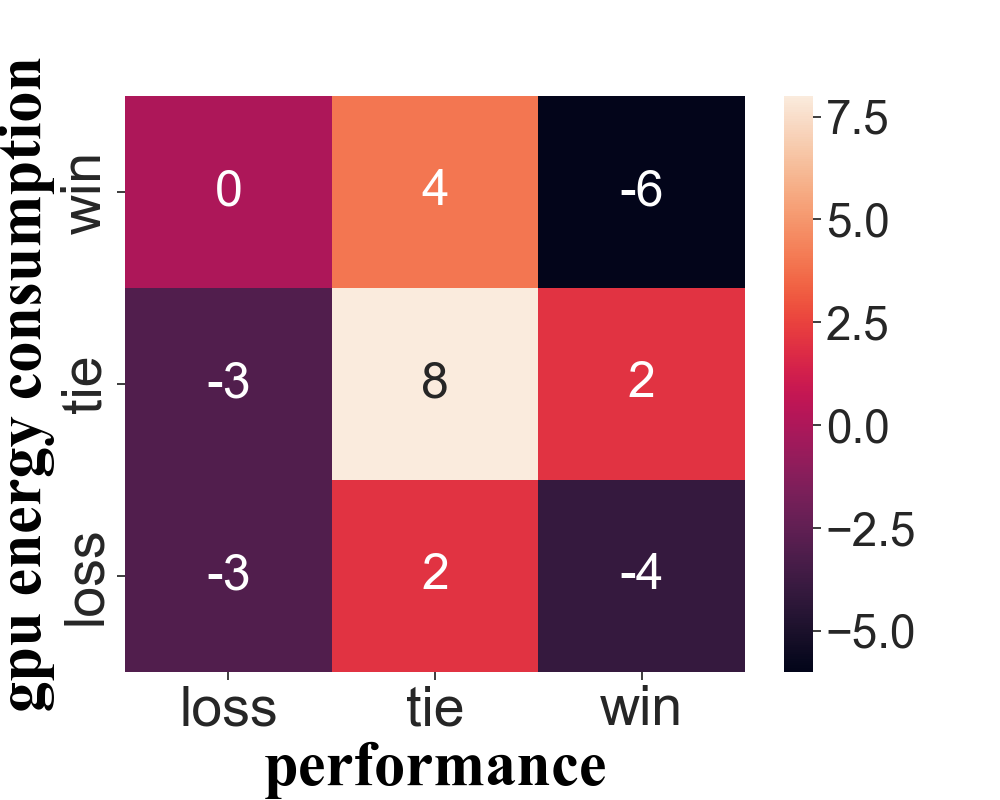}
        (c) gpu energy consumption
    \end{minipage}
    \caption{Difference in the parallel environment while mutating epochs. The numbers in the grids stand for the differences when training in parallel. If the number $> 0$, there are more of this kind of model while training in parallel; 0 means they are the same; the number $< 0$ means there are fewer of this type of model in parallel.}
    \label{fig:rq3-e}
\end{figure}

\begin{figure}[t]
    \centering
    \begin{minipage}{0.32\textwidth}
        \centering
        \includegraphics[scale=0.17]{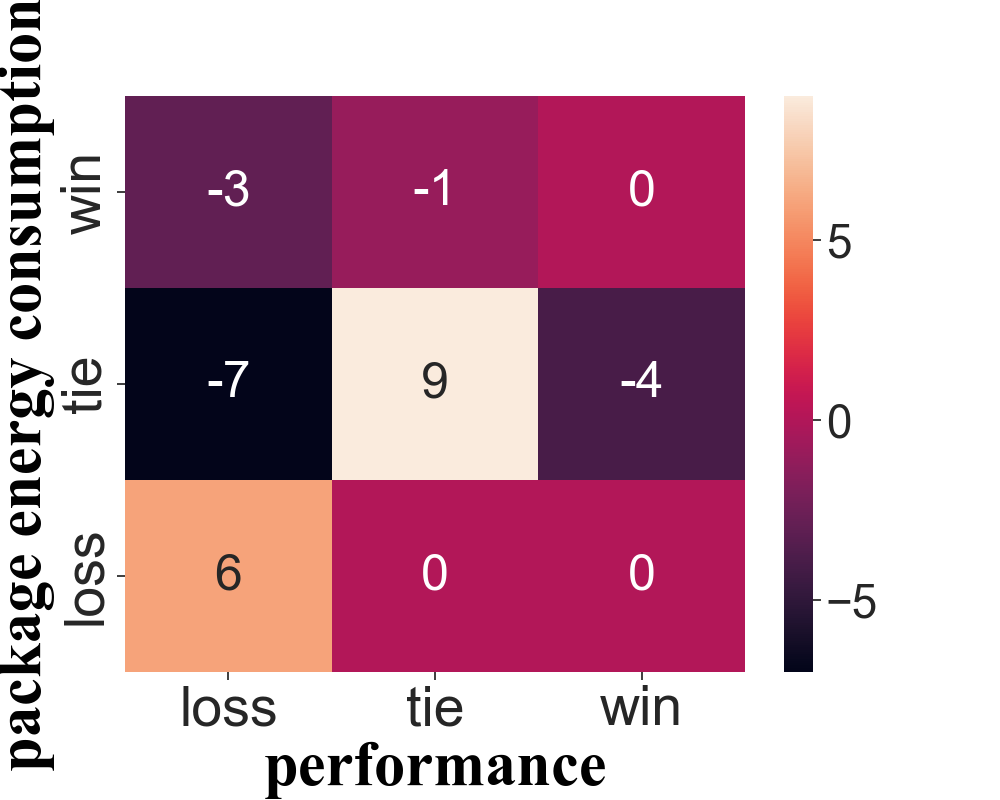}
        (a) package energy consumption
    \end{minipage}
    \begin{minipage}{0.32\textwidth}
        \centering
        \includegraphics[scale=0.17]{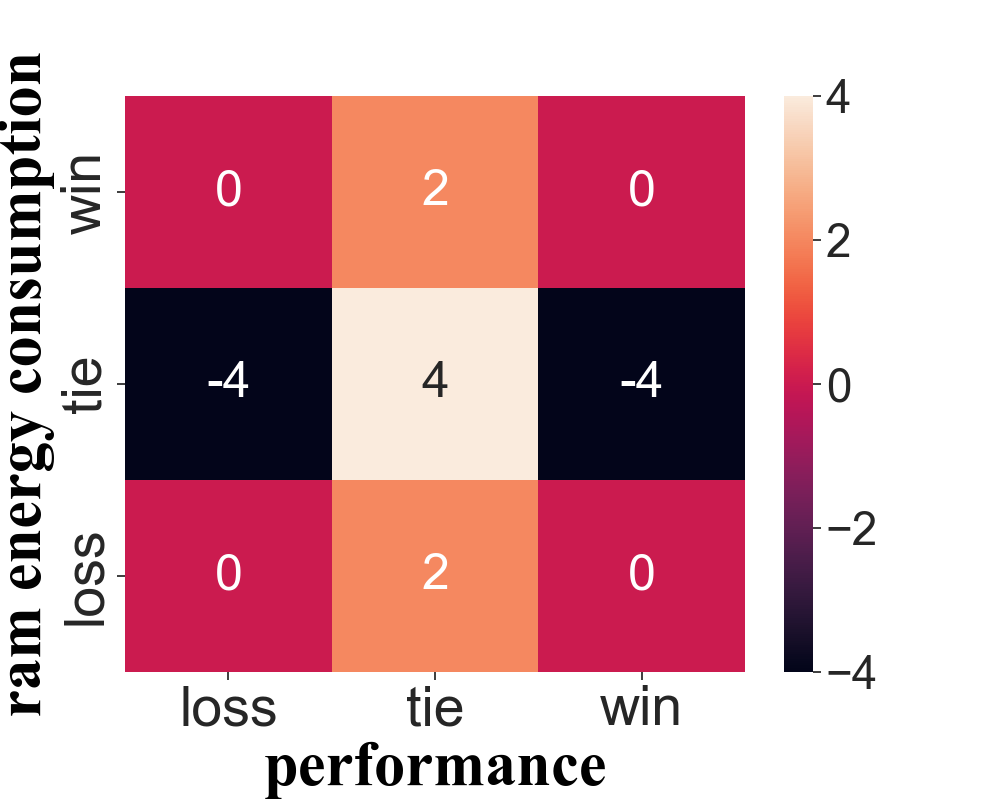}
        (b) ram energy consumption
    \end{minipage}
    \begin{minipage}{0.32\textwidth}
        \centering
        \includegraphics[scale=0.17]{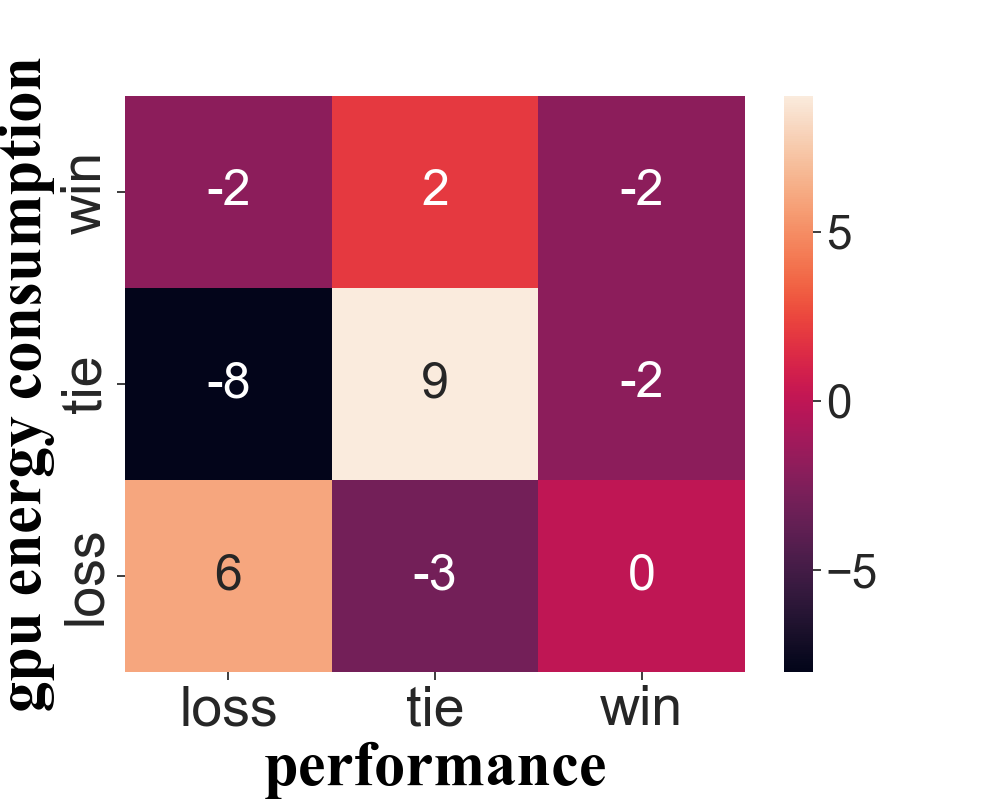}
        (c) gpu energy consumption
    \end{minipage}
    \caption{Difference in the competition situation while mutating learning rate}
    \label{fig:rq3-l}
\end{figure}

\noindent \textbf{Result.}
The results of the parallel analysis are organized in the following tables and figures. Table~\ref{tab:rq3} shows the results of the correlation between hyperparameters and energy metrics in a parallel environment. The trade-off analysis is presented in two figures: Figure~\ref{fig:rq3-e} shows the result between the epochs and energy consumption, and Figure~\ref{fig:rq3-l} presents the analysis between the learning rate and the energy metrics.

Table~\ref{tab:rq3} is the Spearman correlation analysis in the parallel environment. It shows the result in the same way as in Table~\ref{tab:1} in RQ1. From the table, epochs are still significantly positively correlated with energy consumption and time cost. The learning rate has a weak negative relation with the package, GPU, and time cost, which differs from RQ1. As for gamma, it only weakly correlates with RAM in RQ1. However, while training pairs in the parallel environment, adjusting gamma weakly negatively correlates with the package, GPU, and time cost. Therefore, significant correlations in the parallel environment have no difference from RQ1, while weak correlations have slight discrepancies.

\begin{table}[t]
    \begin{center}
        \tabcolsep = 1em
        \caption{The summary of Spearman correlation analysis on the training phase of two models with the other two models running in the background.}
        \label{tab:rq3}
                \small
        \vspace{1ex}
        \begin{tabular}{*{6}{|c}|}
            \hline
            hyperparameter & package & ram   & gpu   & time  & performance\\ \hline  
            \hline      
            epochs         & 4/0/0   & 4/0/0 & 4/0/0 & 4/0/0 & 1/3/0\\ \hline
            learning rate  & 0/3/1   & 0/4/0 & 0/3/1 & 0/3/1 & 3/0/1\\ \hline
            gamma          & 0/3/1   & 2/2/0 & 0/3/1 & 0/4/0 & 1/3/0\\ \hline
        \end{tabular}
    \end{center}
\end{table}

Figures~\ref{fig:rq3-e} and \ref{fig:rq3-l} review the difference of ``win-tie-loss'' in the parallel environment from RQ2. The numbers in each grid show the distinction between common environments and parallel environments. The value is the ``win-tie-loss'' number in parallel times two, minus the sum of the numbers of two models in a common environment. For example, in the last line of Figure~\ref{fig:rq3-e}(a), -3 means there are three fewer models ``lose'' both in energy consumption and performance when training in parallel; 6 means there are six more models ``lose'' in energy consumption and ``tie'' in performance when running in parallel; -1 means there is one fewer model ``loses'' in energy cost and ``wins'' in network performance. Since epochs and the learning rate can be mutated for all pairs, we conduct our analysis on these two hyperparameters.

Figure~\ref{fig:rq3-e} shows the difference when mutating epochs. We can see that training in parallel has more models ``tie'' in performance and fewer model ``loses'' or ``wins'' in energy consumption. This shows that energy consumption is more susceptible to mutated hyperparameters, while performance is more stable in parallel environments. Among the three energy consumption metrics, there are fewer differences in package and RAM, but more distinctions in GPU energy cost. The differences of the package and RAM energy metrics are similar to each other, meaning that these energy consumption are switched in the same direction; while more models ``tie'' to energy consumption and network performance for the GPU energy cost.

Figure~\ref{fig:rq3-l} reveals the distinction when mutating the learning rate. Like mutating epochs, mutating the learning rate leads to more models ``tie'' in performance, but more models ``lose'' in energy consumption. In addition, there are more differences when mutating learning rates than epochs. From the metrics of our models, we find out that model performance is more likely to ``tie'' when training in parallel, which means performance becomes more stable; the energy consumption is more likely to ``win'' or ``lose'', which means energy consumption becomes more sensitive to hyperparameters.

\begin{framed}
        \noindent\textbf{Answer to RQ3.} There are differences when training models in parallel environments. In our observation, energy consumption is more susceptible in parallel environments, while more models ``tie'' in performance. Mutating the learning rate has more disparities from training separately than mutating epochs.
\end{framed}

\section{Discussion}
\label{sec:discussion}
We further discuss the aims and results of our method in the following four aspects: correlation among metrics, distinction among models, case studies, and insights into our work.

\subsection{Correlation among metrics}
Here, we discuss Spearman correlation analysis among metrics. We place Tables~\ref{tab:dis-metrics} and \ref{tab:dis-multi} that are similar to RQ1 (see Table~\ref{tab:1} for details).

Table~\ref{tab:dis-metrics} shows the result of training not in parallel. %Table~\ref {tab:dis-multi} shows the correlation analysis result in our parallel environment. 
Compared to Table~\ref{tab:1} in RQ1, we sum all hyperparameter mutations and perform the Spearman correlation analysis on every pair of two metrics. For example, ``2/12/1'' in Table~\ref{tab:dis-metrics} means two samples show positive correlations; 12 samples have no significant correlation; one sample has a negative correlation. As can be seen, 
all three energy consumption metrics and time cost have a weak correlation with performance. Time cost is strongly related to energy consumption. Also, the three energy consumption metrics have a significant correlation with each other.

\begin{table}[t]
    \begin{center}
        \tabcolsep = 1em
        \small
        \caption{Spearman correlation analysis among metrics}
        \label{tab:dis-metrics}
        \vspace{1ex}
        \begin{tabular}{*{5}{|c}|}
            \hline
                        & package & ram   & gpu   & time\\ \hline    
            \hline    
            performance & 2/12/1   & 2/13/0 & 1/14/0 & 3/12/0\\ \hline
            time         & 14/1/0   & 15/0/0 & 14/1/0 & -\\ \hline
            gpu         & 13/2/0   & 11/4/0 &   -   & -\\ \hline
            ram         & 11/4/0   &   -   &   -   & -\\ \hline
        \end{tabular}
    \end{center}
\end{table}

Table~\ref{tab:dis-multi} indicates the correlations in the parallel environment. In the parallel environment, energy consumption metrics and time cost also have a weak correlation with performance. However, the GPU has a weaker relationship with time cost and RAM. Other energy metrics are significantly related to each other. From the two tables, we can discover that very few samples show negative correlations in both environments.

\begin{table}[t]
    \begin{center}
        \tabcolsep = 1em
        \small
        \caption{Spearman correlation analysis among metrics in the parallel environment}
        \label{tab:dis-multi}
        \vspace{1ex}
        \begin{tabular}{*{5}{|c}|}
            \hline
                        & package & ram   & gpu   & time\\ \hline  \hline  
            performance & 2/10/0   & 2/10/0 & 3/9/0 & 2/10/0\\ \hline
            time        & 11/1/0   & 12/0/0 & 8/4/0 & -\\ \hline
            gpu         & 12/0/0   & 9/3/0 &   -   & -\\ \hline
            ram         & 10/2/0   &   -   &   -   & -\\ \hline
        \end{tabular}
    \end{center}
\end{table}

\subsection{Distinction among models}
In this section, we will discuss the differences among models. When we are testing on energy consumption, we notice that there are also distinctions among models. Hence, we list the details of the correlation analysis for each model instead of each hyperparameter in Tables~\ref{tab:dis-e} and \ref{tab:dis-lr}.

Table~\ref{tab:dis-e} shows the correlation of five models when mutating epochs. From the table, we can see that there is nearly no difference when mutating epochs. Therefore, in all models, the energy cost and performance of models almost react in the same way when mutating epochs.

\begin{table}[t]
    \begin{center}
        \tabcolsep = 1em
        \caption{The summary of Spearman correlation analysis on the epochs of five models.}
        \label{tab:dis-e}
        \vspace{1ex}
        \small
        \begin{tabular}{*{6}{|c}|}
            \hline
            model & package & ram   & gpu   & time  & performance\\ \hline        \hline
            $\mathcal{A}$         & 1   & 1 & 1 & 1 & 0\\ \hline
            $\mathcal{B}$         & 1   & 1 & 1 & 1 & 0\\ \hline
            $\mathcal{C}$         & 1   & 1 & 0 & 1 & 0\\ \hline
            $\mathcal{D}$         & 1   & 1 & 1 & 1 & 0\\ \hline
            $\mathcal{E}$         & 1   & 1 & 1 & 1 & 0\\ \hline
        \end{tabular}
        \begin{tablenotes}
            \footnotesize
            \item The table shows the result of the correlation analysis. ``1'' means the Spearman correlation coefficient $\ge 0.2$ and the $p$-value $< 0.05$. ``-1'' means the Spearman correlation coefficients $\le -0.2$ and $p$-value $< 0.05$. The other situation belongs to ``0''.
        \end{tablenotes}
    \end{center}
\end{table}

Table~\ref{tab:dis-lr} shows the distinctions among five models when mutating the learning rate. There is a significant distinction in network performance, with one positive and two negative correlations. The last two models with larger datasets have a negative correlation with our models. As for energy consumption, there are only negative correlations. However, only RAM, GPU, and run-time negatively correlate with the learning rate in some models. From the tables, there are distinctions among the models. However, it still depends on the hyperparameters mutated.

\begin{table}[t]
    \begin{center}
        \tabcolsep = 1em
        \caption{The summary of Spearman correlation analysis on the learning rate of five models.}
        \label{tab:dis-lr}
        \vspace{1ex}
        \small
        \begin{tabular}{*{6}{|c}|}
            \hline   
            model & package & ram   & gpu   & time  & performance\\ \hline     \hline     
            $\mathcal{A}$        & 0 & 0  & 0  & 0  & 0\\ \hline
            $\mathcal{B}$        & 0 & -1 & -1 & -1 & 1\\ \hline
            $\mathcal{C}$        & 0 & 0  & -1 & 0  & 0\\ \hline
            $\mathcal{D}$        & 0 & 0  & -1 & 0  & -1\\ \hline
            $\mathcal{E}$        & 0 & 0  & 0  & 0  & -1\\ \hline
        \end{tabular}
    \end{center}
\end{table}

\subsection{Case studies}
We present some case studies in this section to further explain our conclusions. These cases are picked from former experiments to help understand our answers for RQs.

\subsubsection{An example for correlation analysis between energy consumption and epochs}
\begin{figure}[t]
    \centering
    \includegraphics[scale=0.5]{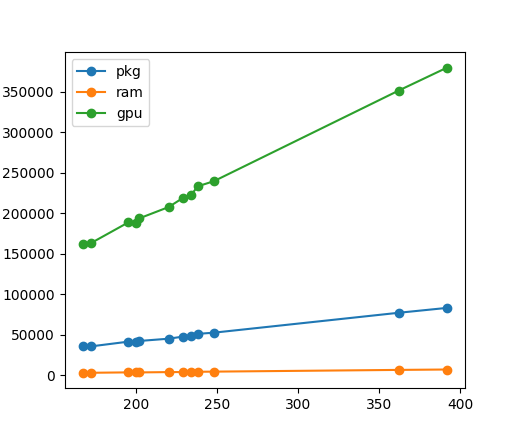}
    \caption{The relation between energy consumption and epochs on Model $\mathcal{E}$. The x-axis shows the value of mutated epochs, and the y-axis shows the value of energy consumption.}
    \label{fig:dis-case}
\end{figure}

Figure~\ref{fig:dis-case} shows the details of the energy consumption of mutated epochs in Model $\mathcal{E}$. From the figure, we can observe that the epochs are significantly correlative with three energy metrics. In RQ1, we find that hyperparameters have correlations with energy consumption. The three lines in the figure support the result we found in RQ1. We can see significant positive correlations between epochs and three energy cost metrics.

\subsubsection{Examples for a trade-off between energy consumption and network performance}
In RQ2, we reveal that mutating epochs consume less energy and mutating learning rate ``wins'' in network performance. For GPU, mutating the weight decay and the learning rate can lead to less energy cost. Here we list part of the experiment result of RQ2 to discuss this conclusion.

Table~\ref{tab:dis-tf} shows examples that can support this conclusion. From the table, when epochs are 72, which is greater than the default value (60), all metrics of energy consumption are significantly greater than the original value. Therefore, decreasing the epochs can gain a greener model. When the learning rate (0.099) is greater than the default value (0.05), the GPU energy consumption and performance value are significantly less than the original values. So, reducing the learning rate here can make the model greener. When weight decay (0.0009) is greater than its default value (0.0005), the GPU energy consumption is less than its original one. Therefore, we can decline the weight decay for a greener model. They all support the conclusion we found in RQ2.

\begin{table}[t]
    \begin{center}
        \tabcolsep = 1em
        \caption{The difference between three models and the original model.}
        \small
        \label{tab:dis-tf}
        \vspace{1ex}
        \begin{tabular}{*{6}{|c}|}
            \hline   
            hyperparameter & package & ram   & gpu   & time  & performance\\ \hline\hline          
            epochs (72)            & 1 & 1  & 1  & 1  & 0\\ \hline
            learning rate (0.099)  & 0 & 0  & -1 & 0  & -1\\ \hline
            weight decay (0.0009)  & 0 & 0  & -1 & 0  & 0\\ \hline
        \end{tabular}
        \begin{tablenotes}
            \footnotesize
            \item The table shows three models from the Model $\mathcal{E}$. The value in the bracket is the mutated value of hyperparameters. ``1'' means the value is significantly greater than the original ($p < 0.05$ and $\delta \ge 0.147$). ``-1'' means the value is significantly lower than the original ($p < 0.05$ and $\delta \le -0.147$). The situation left is ``0'' ($p \ge 0.05$ or $-0.147 \le \delta \le 0.147$). 
        \end{tablenotes}
    \end{center}
\end{table}

\subsubsection{An example for the difference in the parallel environment}
In the parallel environment, we train a model in the background, then train another model, and monitor the total energy consumption and performance of the model we monitor.

\begin{figure}[t]
    \centering
    \begin{minipage}{0.32\textwidth}
        \centering
        \includegraphics[scale=0.17]{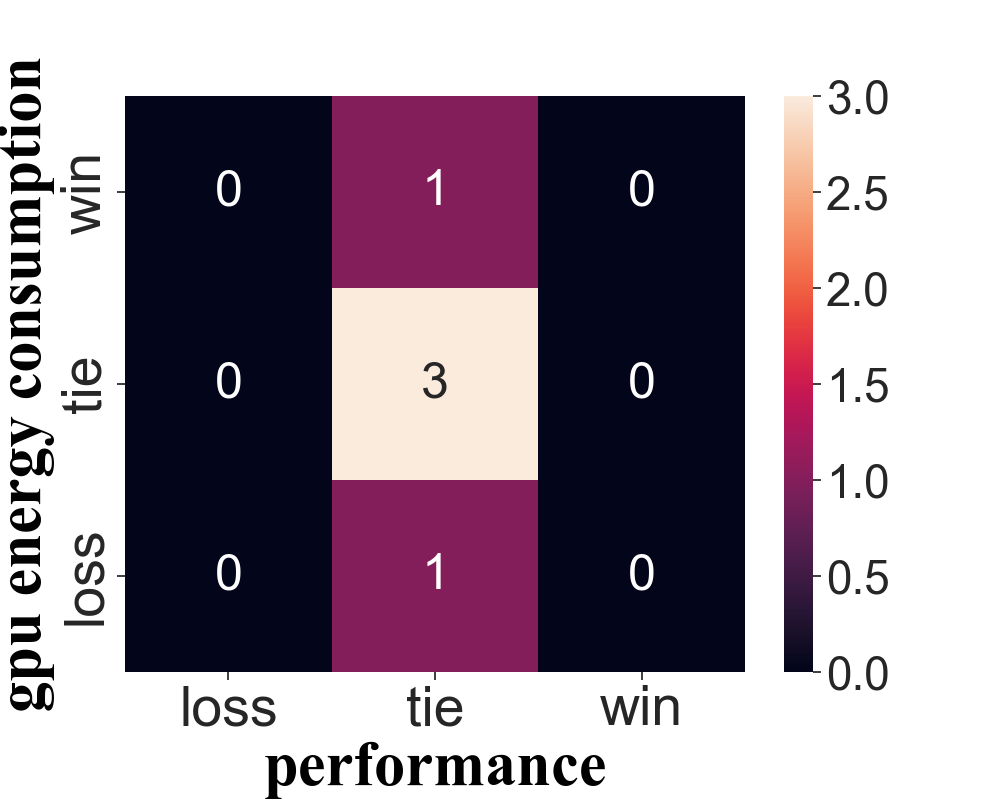}
        (a) Model $\mathcal{A}$ trained and monitored with Model $\mathcal{C}$
    \end{minipage}
    \begin{minipage}{0.32\textwidth}
        \centering
        \includegraphics[scale=0.17]{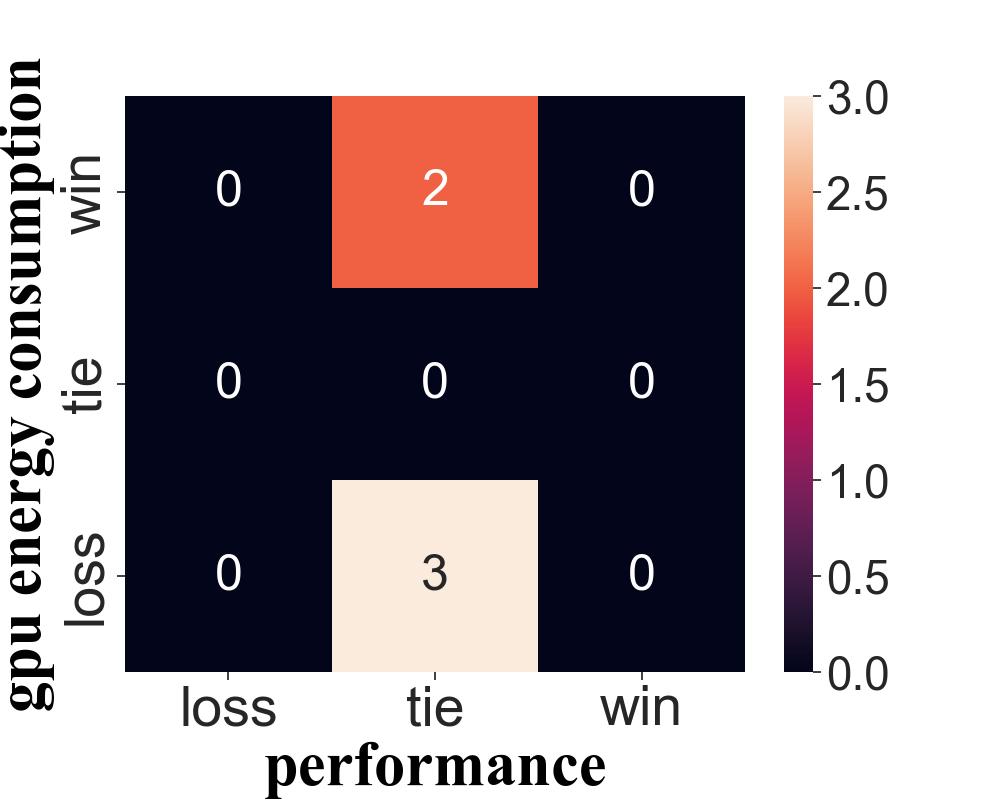}
        (b) Model $\mathcal{A}$ trained separately
    \end{minipage}
    \begin{minipage}{0.32\textwidth}
        \centering
        \includegraphics[scale=0.17]{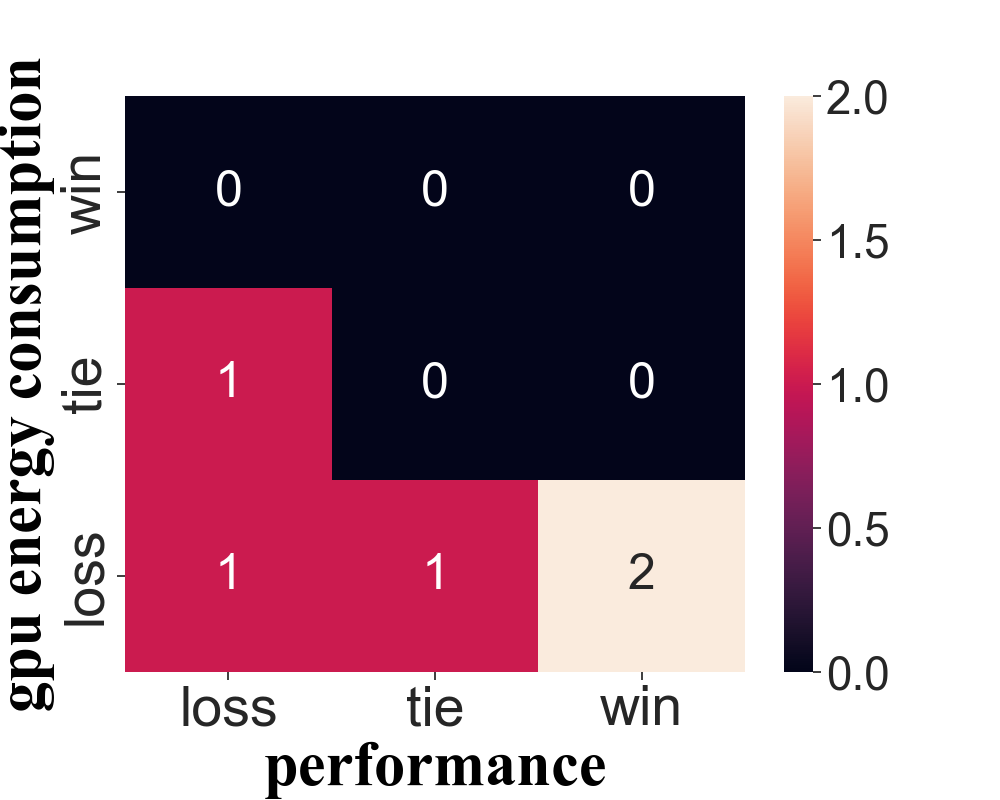}
        (c) Model $\mathcal{C}$ trained separately
    \end{minipage}
    \caption{Difference in the competition situation of GPU consumption while mutating epoch}
    \label{fig:dis-rq3}
\end{figure}

We show an example in Figure~\ref{fig:dis-rq3} to show the difference in a parallel environment. The example contains training two models not in parallel and training them in a parallel environment. The heat maps (the approach to reading the figure can be found in Figure~\ref{fig:rq3-e} in RQ3) in the figure show the trade-off between GPU energy cost and network performance when mutating epochs. The results are organized by the three situations (i.e., train and monitor Model $\mathcal{A}$ with Model $\mathcal{C}$ and train two models separately). Like the conclusion we make in RQ3, more models ``win'' or ``tie'' in energy metrics in parallel, which proves that energy consumption is easier to change by hyperparameters. In addition, models ``tie'' in performance when two models train in a parallel environment. This shows that models are more stable in a parallel environment.

\subsection{Insight of our work}
In this section, we will introduce insights into our work and what they bring to DL programmers or further research.

\subsubsection{Help to develop greener models}
In RQ1, we find that hyperparameters are related to the energy consumption of the model training phase. In addition, in RQ2, we uncover models with better trade-offs that consume less energy without harming network performance or train a better network without using more energy. For most models we tested, slightly declining epochs in a small range can train the network in a greener way, while a carefully adjusted learning rate can make the training phase greener for specific models, especially the models that consume mainly GPU energy. 

Therefore, practitioners of DL models can pay more attention to the energy consumption of the models while adjusting hyperparameters. However, some questions are still unanswered; for instance, we can see that the correlations differ for different models. Hence, further studies can focus on the green DL models.

\subsubsection{Reveal the Difference for parallel environments}
Our research indicates that mutating different hyperparameters in parallel environments has a distinction from training separately. Therefore, the further details of parallel environments are still worth investigating. For instance, the amount and kind of models training in parallel may lead to different energy consumption results. In further studies, we can pay more attention to parallel environments to solve the energy issues.

\section{Threats to validity}
\label{sec:threats}
In this section, we list the main threats to the validity of our work in the following six aspects.

Our work depends on the hardware, which means the result may have distinctions on different hardware. Since the tools we used are related to CPU and GPU structure, the result differs among devices. Especially on some servers, the tools we used could not even collect the power metrics. However, our objective is to discover the relationship between hyperparameters and energy consumption. Therefore, hardware differences are ignored in this work. Studies on how devices influence energy consumption can be carried out in the future.

There are some accuracy deviations in metrics collection. Perf can only measure the total energy consumption of the whole device. Nvidia can only measure the total energy consumption of a single GPU core. Therefore, noise, like background processes, can affect metrics collection. This issue also occurs in other works, which is hard to neutralize. In our experiments, we shut down as many processes and set idle times between training to diminish the accuracy deviations in metrics collection.

The models we selected affected the results we got. Due to the mutation approach we developed, many models are not selected (the rule of model selection can be read in Section~\ref{sec:appraoch}). These models usually do not have proper performance metrics, such as GAN~\cite{gan}. Also, due to the limitation of our server and numerous repeated experiments demanded by the mutation test, only models with a sufficiently short training phase are chosen.

The mutation range of hyperparameters makes a difference in our results. In this research, we mutate them in a small enough range that does not harm the network performance. However, when we test the range, we observe that the relations between hyperparameters and energy consumption cause distinctions in the correlations. However, our purpose is to imitate how practitioners and programmers adjust the network. So we finally limit the range to a reasonable and small enough one.

In RQ3, we test the parallel environment. However, we only test the parallel environment in which two models are trained in parallel because of our limitation of computing resources. Since our results show differences between training one model and training two models in parallel, training models of more than two may also have differences. We plan to finish how energy consumption shifts when multiple models are trained in parallel environments with more computing resources in the future.

Randomness is also a threat to the validity of our research. Considering the limitation of our computing resources, we finally chose to repeat each mutant five times to reduce the influence of randomness.

\section{Related work}
\label{sec:related}\
This section introduces the work related to this paper from three points of view. They are green software and artificial intelligence, DL testing and debugging, and mutation testing.
\subsection{Green software and artificial intelligence}
Many works focus on the energy consumption of software, which leads to green software~\cite{9585139}. Fu et al.~\cite{Estimating-Software-Energy} conducted a model of energy consumption characterized by performance events to explain the origin of energy consumption with high accuracy. An efficient approach is outlined by Mehra et al.~\cite{Towards-a-Green} to measure the ``greenness'' of software according to the different choices in software development lifecycle dimensions. Kumar et al.~\cite{Energy-Efficient-Machine} presented Java Energy Profiler \& Optimizer, an Eclipse plugin that can profile and optimize the source code of ML software. Warade et al.~\cite{Optimising-workflow-execution} proposed a generic framework that develops an optimal schedule by user constraints, performance, and energy consumption factors for workflow execution. Weber et al.~\cite{Twins-or-False} conducted an empirical study on tools and methods that directly reduce energy consumption. Sarro et al.~\cite{Search-Based-Software} found that Information and Communication Technologies increase the carbon footprint and use Search-Based Software Engineering to produce greener software.

Some of these works contributed to smartphones and embedded systems. Mcintosh et al.~\cite{What-can-Android} concluded that some DL algorithms cost less energy and perform better. Bangash et al.~\cite{Black-Box-Technique} found that byte-code transformations can reduce the energy consumption of smartphone applications. Bangash.~\cite{Cost-effective-Strategies} developed a general methodology to extract energy-efficient guidelines that help energy-efficient application developments.

With more software using DL technology, several studies on energy consumption on DL. Georgiou et al.~\cite{greenAi} conducted an evaluation that raised awareness of the energy costs of different DL frameworks. Zhang et al.~\cite{pCAMP:-Performance-Comparison} made a comparison of state-of-the-art ML packages in terms of latency, memory footprint, and energy consumption. Garca-Martn et al.~\cite{Estimation-of-energy} estimated the energy consumption of the different ML approaches. Haque et al.~\cite{EREBA-Black-box} proposed EREBA, a black-box testing method for an adaptive neural network that determines energy robustness.

\subsection{DL testing and debugging}
Software today is increasing the use of DL components, which leads to the increasing importance of DL testing and debugging. A white-box framework is proposed by Pei et al.~\cite{DeepXplore-Automated-Whitebox} to test real-world DL systems that can detect erroneous behaviors. Ma et al.~\cite{MODE-automated-neural} conducted a model debugging technique that fixes model bugs without introducing new bugs effectively and efficiently. For the robustness of DL systems, Ma et al.~\cite{DeepCT-Tomographic-Combinatorial} performed an exploratory study with combinatorial testing. Gerasimou et al.~\cite{Importance-Driven-Deep} introduced DeepImportance, a systematic testing methodology accompanied by an Importance-Driven test criterion. Zhang et al.~\cite{AUTOTRAINER-An-Automatic} proposed AUTOTRAINER, a tool to detect and repair training problems for DNNs. Nigenda et al.~\cite{Amazon-SageMaker-Model} conducted Amazon SageMaker Model Monitor, a managed service designed for continuous monitoring. To find out how experts and novices debug and train models, Schoop et al.~\cite{UMLAUT-Debugging-Deep} presented Umlaut to check the structure and behavior of DL models.

For recent works, Gao et al.~\cite{Metamorphic-Testing-of} proposed a method for machine translation testing using back-translation. Hu et al.~\cite{Aries-Efficient-Testing} presented Aries to estimate the performance of DNNs on new and unlabeled data. Monjezi et al.~\cite{Information-Theoretic-Testing} designed Dice, an information-theoretic framework that discovers and localizes DNN fairness defects. The cost-effectiveness of composite metamorphic relations is measured by Arrieta~\cite{On-the-Cost} for testing DL systems. Gao et al.~\cite{Adaptive-Test-Selection} presented ATS, an adaptive test selection method, which deals with the problems caused by unlabelled datasets. To explore the target library of DL, Gu et al.~\cite{Muffin-testing-deep} proposed Muffin to generate various DL models. Wang et al.~\cite{EAGLE-Creating-Equivalent} proposed EAGLE that uses equivalent graphs to conduct differential testing on the implementation of DL.

\subsection{Traditional and DL Mutation Testing}
Mutation test is a traditional software testing technique that has been used for more than four decades to measure the adequacy of tests~\cite{mutation_test}, which is widely used. For example, to locate the causes of multilingual bugs in the real world, Hong et al.~\cite{HONG} introduced MUSEUM, a mutation-based fault localization technique. With the spread of the Android system, Deng et al.~\cite{android} applied mutation tests to Android applications. Wu et al.~\cite{mem} focused on addressing memory failures and tested 18 open-source programs, including two large real-world programs. For aspect-oriented modeling, at the aspect-oriented model level, Lindstr\"{o}m et al.~\cite{aom} presented a method to design abstract tests.

With the introduction of new technologies, more tests focus on these new technologies, like DL and quantum computing. For cyber-physical systems, Cornejo et al.~\cite{cps} proposed MASS, a mutation analysis tool for embedded software. Building on the definition of syntactically equivalent quantum operations, Fortunato et al.~\cite{quantum} investigated mutation operators based on quantum gates and qubit measurements. For ML software systems, Laurent et al.~\cite{jsimutate} conducted JSIMUTATE that analyzes the performance of models. With more mutation test tools relying on DL models, Ojdanic et al.~\cite{mutateion_tools} proposed criteria to compare these mutation tools.

Besides, several works contribute to mutation usefulness and efficiency. Predictive Mutation Testing (PMT) is proposed by Zhang et al.~\cite{pmt} to predict the results of mutation testing without executing mutants. To refine the test suite, Delgado-Pérez et al.~\cite{Search-based} used Evolutionary Mutation Testing (EMT) to preserve energy consumption by reducing the number of mutants. Just et al.~\cite{inferring} performed a selective mutation strategy that reduces the inherent redundancy of performing a full mutation analysis at little cost to obtain most of its benefits. Tian et al.~\cite{leam} proposed LEAM, a DL-based mutation technique, to construct high-quality mutation faults for DL base software.

\section{Conclusion}
\label{sec:conclusion}
With the widespread use of DL software, green DL has become a vitally important issue, which can diminish carbon dioxide production and software financial costs. This paper contributes to the effects on the hardware energy consumption from DL hyperparameters in both the common and parallel environments. We employ mutation operators to collect the energy consumption metrics of different hyperparameters. Based on our evaluation, we report the analysis results with these data.

We observe that most hyperparameters have a positive or negative correlation with hardware energy consumption. In addition, making the DL models greener is feasible by adjusting the hyperparameters. Also, in a parallel environment, the relationship between hyperparameters and energy consumption is different; from our samples, energy consumption becomes easier to affect in parallel. Finally, this paper has the following insights for researchers and practitioners. First, since the energy consumption of model training can be reduced by adjusting the DL hyperparameters, DL developers should pay more attention to the energy consumption of the models while adjusting hyperparameters. Second, DL developers should pay attention to the environment, whether models are trained with other models in the parallel environment; the training environments also influence the relationship between hyperparameters and energy consumption.

\section*{Replication Package}
The source code and data for conducting this study are presented at \textbf{\url{https://github.com/IIllIlllIl/greenAi.git}}.

\bibliographystyle{elsarticle-num}
\bibliography{greenai}

@ARTICLE{9585139,
  author={Verdecchia, Roberto and Lago, Patricia and Ebert, Christof and de Vries, Carol},
  journal={IEEE Software}, 
  title={Green IT and Green Software}, 
  year={2021},
  volume={38},
  number={6},
  pages={7-15},
  doi={10.1109/MS.2021.3102254}}

@inproceedings{humbatova2021deepcrime,
  title={Deepcrime: mutation testing of deep learning systems based on real faults},
  author={Humbatova, Nargiz and Jahangirova, Gunel and Tonella, Paolo},
  booktitle={Proceedings of the 30th ACM SIGSOFT International Symposium on Software Testing and Analysis},
  pages={67--78},
  year={2021}
}

@article{shen2021boundary,
  title={Boundary sampling to boost mutation testing for deep learning models},
  author={Shen, Weijun and Li, Yanhui and Han, Yuanlei and Chen, Lin and Wu, Di and Zhou, Yuming and Xu, Baowen},
  journal={Information and Software Technology},
  volume={130},
  pages={106413},
  year={2021},
  publisher={Elsevier}
}

@article{li2022higher,
  title={How higher order mutant testing performs for deep learning models: A fine-grained evaluation of test effectiveness and efficiency improved from second-order mutant-classification tuples},
  author={Li, Yanhui and Shen, Weijun and Wu, Tengchao and Chen, Lin and Wu, Di and Zhou, Yuming and Xu, Baowen},
  journal={Information and Software Technology},
  volume={150},
  pages={106954},
  year={2022},
  publisher={Elsevier}
}

@inproceedings{hu2019deepmutation++,
  title={Deepmutation++: A mutation testing framework for deep learning systems},
  author={Hu, Qiang and Ma, Lei and Xie, Xiaofei and Yu, Bing and Liu, Yang and Zhao, Jianjun},
  booktitle={2019 34th IEEE/ACM International Conference on Automated Software Engineering (ASE)},
  pages={1158--1161},
  year={2019},
  organization={IEEE}
}

@inproceedings{ma2018deepmutation,
  title={Deepmutation: Mutation testing of deep learning systems},
  author={Ma, Lei and Zhang, Fuyuan and Sun, Jiyuan and Xue, Minhui and Li, Bo and Juefei-Xu, Felix and Xie, Chao and Li, Li and Liu, Yang and Zhao, Jianjun and others},
  booktitle={2018 IEEE 29th international symposium on software reliability engineering (ISSRE)},
  pages={100--111},
  year={2018},
  organization={IEEE}
}

@inproceedings{shen2018munn,
  title={Munn: Mutation analysis of neural networks},
  author={Shen, Weijun and Wan, Jun and Chen, Zhenyu},
  booktitle={2018 IEEE international conference on software quality, reliability and security companion (QRS-C)},
  pages={108--115},
  year={2018},
  organization={IEEE}
}

@article{chen2011overview,
  title={An overview of energy consumption of the globalized world economy},
  author={Chen, Zhan-Ming and Chen, GQ},
  journal={Energy Policy},
  volume={39},
  number={10},
  pages={5920--5928},
  year={2011},
  publisher={Elsevier}
}

@inproceedings{chen2021empirical,
  title={An empirical study on deployment faults of deep learning based mobile applications},
  author={Chen, Zhenpeng and Yao, Huihan and Lou, Yiling and Cao, Yanbin and Liu, Yuanqiang and Wang, Haoyu and Liu, Xuanzhe},
  booktitle={2021 IEEE/ACM 43rd International Conference on Software Engineering (ICSE)},
  pages={674--685},
  year={2021},
  organization={IEEE}
}

@inproceedings{zhang2019empirical,
  title={An empirical study of common challenges in developing deep learning applications},
  author={Zhang, Tianyi and Gao, Cuiyun and Ma, Lei and Lyu, Michael and Kim, Miryung},
  booktitle={2019 IEEE 30th International Symposium on Software Reliability Engineering (ISSRE)},
  pages={104--115},
  year={2019},
  organization={IEEE}
}

@inproceedings{zhang2020empirical,
  title={An empirical study on program failures of deep learning jobs},
  author={Zhang, Ru and Xiao, Wencong and Zhang, Hongyu and Liu, Yu and Lin, Haoxiang and Yang, Mao},
  booktitle={Proceedings of the ACM/IEEE 42nd International Conference on Software Engineering},
  pages={1159--1170},
  year={2020}
}

@article{li2023survey,
  title={Survey on evolutionary deep learning: Principles, algorithms, applications, and open issues},
  author={Li, Nan and Ma, Lianbo and Yu, Guo and Xue, Bing and Zhang, Mengjie and Jin, Yaochu},
  journal={ACM Computing Surveys},
  volume={56},
  number={2},
  pages={1--34},
  year={2023},
  publisher={ACM New York, NY}
}

@article{hertel2020sherpa,
  title={Sherpa: Robust hyperparameter optimization for machine learning},
  author={Hertel, Lars and Collado, Julian and Sadowski, Peter and Ott, Jordan and Baldi, Pierre},
  journal={SoftwareX},
  volume={12},
  pages={100591},
  year={2020},
  publisher={Elsevier}
}

@inproceedings{akiba2019optuna,
  title={Optuna: A next-generation hyperparameter optimization framework},
  author={Akiba, Takuya and Sano, Shotaro and Yanase, Toshihiko and Ohta, Takeru and Koyama, Masanori},
  booktitle={Proceedings of the 25th ACM SIGKDD international conference on knowledge discovery \& data mining},
  pages={2623--2631},
  year={2019}
}

@article{tong2020economic,
  title={Economic growth, energy consumption, and carbon dioxide emissions in the E7 countries: a bootstrap ARDL bound test},
  author={Tong, Teng and Ortiz, Jaime and Xu, Chuanhua and Li, Fangjhy},
  journal={Energy, Sustainability and Society},
  volume={10},
  number={1},
  pages={1--17},
  year={2020},
  publisher={Springer}
}

@inproceedings{DBLP:conf/issre/0001GMLK19,
  author       = {Tianyi Zhang and
                  Cuiyun Gao and
                  Lei Ma and
                  Michael R. Lyu and
                  Miryung Kim},
  editor       = {Katinka Wolter and
                  Ina Schieferdecker and
                  Barbara Gallina and
                  Michel Cukier and
                  Roberto Natella and
                  Naghmeh Ramezani Ivaki and
                  Nuno Laranjeiro},
  title        = {An Empirical Study of Common Challenges in Developing Deep Learning
                  Applications},
  booktitle    = {30th {IEEE} International Symposium on Software Reliability Engineering,
                  {ISSRE} 2019, Berlin, Germany, October 28-31, 2019},
  pages        = {104--115},
  publisher    = {{IEEE}},
  year         = {2019},
  url          = {https://doi.org/10.1109/ISSRE.2019.00020},
  doi          = {10.1109/ISSRE.2019.00020},
  bibsource    = {dblp computer science bibliography, https://dblp.org}
}

@inproceedings{DBLP:conf/icse/GoharBR23,
  author       = {Usman Gohar and
                  Sumon Biswas and
                  Hridesh Rajan},
  title        = {Towards Understanding Fairness and its Composition in Ensemble Machine
                  Learning},
  booktitle    = {45th {IEEE/ACM} International Conference on Software Engineering,
                  {ICSE} 2023, Melbourne, Australia, May 14-20, 2023},
  pages        = {1533--1545},
  publisher    = {{IEEE}},
  year         = {2023},
  url          = {https://doi.org/10.1109/ICSE48619.2023.00133},
  doi          = {10.1109/ICSE48619.2023.00133},
  bibsource    = {dblp computer science bibliography, https://dblp.org}
}

@inproceedings{DBLP:conf/sigsoft/BiswasR20,
  author       = {Sumon Biswas and
                  Hridesh Rajan},
  editor       = {Prem Devanbu and
                  Myra B. Cohen and
                  Thomas Zimmermann},
  title        = {Do the machine learning models on a crowd sourced platform exhibit
                  bias? an empirical study on model fairness},
  booktitle    = {{ESEC/FSE} '20: 28th {ACM} Joint European Software Engineering Conference
                  and Symposium on the Foundations of Software Engineering, Virtual
                  Event, USA, November 8-13, 2020},
  pages        = {642--653},
  publisher    = {{ACM}},
  year         = {2020},
  url          = {https://doi.org/10.1145/3368089.3409704},
  doi          = {10.1145/3368089.3409704},
  bibsource    = {dblp computer science bibliography, https://dblp.org}
}

@article{DBLP:journals/tosem/HuGCXMPT22,
  author       = {Qiang Hu and
                  Yuejun Guo and
                  Maxime Cordy and
                  Xiaofei Xie and
                  Lei Ma and
                  Mike Papadakis and
                  Yves Le Traon},
  title        = {An Empirical Study on Data Distribution-Aware Test Selection for Deep
                  Learning Enhancement},
  journal      = {{ACM} Trans. Softw. Eng. Methodol.},
  volume       = {31},
  number       = {4},
  pages        = {78:1--78:30},
  year         = {2022},
  url          = {https://doi.org/10.1145/3511598},
  doi          = {10.1145/3511598},
  bibsource    = {dblp computer science bibliography, https://dblp.org}
}

@inproceedings{DBLP:conf/sigsoft/YanTLZMX020,
  author       = {Shenao Yan and
                  Guanhong Tao and
                  Xuwei Liu and
                  Juan Zhai and
                  Shiqing Ma and
                  Lei Xu and
                  Xiangyu Zhang},
  editor       = {Prem Devanbu and
                  Myra B. Cohen and
                  Thomas Zimmermann},
  title        = {Correlations between deep neural network model coverage criteria and
                  model quality},
  booktitle    = {{ESEC/FSE} '20: 28th {ACM} Joint European Software Engineering Conference
                  and Symposium on the Foundations of Software Engineering, Virtual
                  Event, USA, November 8-13, 2020},
  pages        = {775--787},
  publisher    = {{ACM}},
  year         = {2020},
  url          = {https://doi.org/10.1145/3368089.3409671},
  doi          = {10.1145/3368089.3409671},
  bibsource    = {dblp computer science bibliography, https://dblp.org}
}

@article{zhang2021cagfuzz,
  title={Cagfuzz: coverage-guided adversarial generative fuzzing testing for image-based deep learning systems},
  author={Zhang, Pengcheng and Ren, Bin and Dong, Hai and Dai, Qiyin},
  journal={IEEE Transactions on Software Engineering},
  volume={48},
  number={11},
  pages={4630--4646},
  year={2021},
  publisher={IEEE}
}

@inproceedings{DBLP:conf/issta/WangLHCZZ23,
  author       = {Jun Wang and
                  Yanhui Li and
                  Xiang Huang and
                  Lin Chen and
                  Xiaofang Zhang and
                  Yuming Zhou},
  editor       = {Ren{\'{e}} Just and
                  Gordon Fraser},
  title        = {Back Deduction Based Testing for Word Sense Disambiguation Ability
                  of Machine Translation Systems},
  booktitle    = {Proceedings of the 32nd {ACM} {SIGSOFT} International Symposium on
                  Software Testing and Analysis, {ISSTA} 2023, Seattle, WA, USA, July
                  17-21, 2023},
  pages        = {601--613},
  publisher    = {{ACM}},
  year         = {2023},
  url          = {https://doi.org/10.1145/3597926.3598081},
  doi          = {10.1145/3597926.3598081},
  bibsource    = {dblp computer science bibliography, https://dblp.org}
}

@inproceedings{DBLP:conf/sosp/PeiCYJ17,
  author       = {Kexin Pei and
                  Yinzhi Cao and
                  Junfeng Yang and
                  Suman Jana},
  title        = {DeepXplore: Automated Whitebox Testing of Deep Learning Systems},
  booktitle    = {Proceedings of the 26th Symposium on Operating Systems Principles,
                  Shanghai, China, October 28-31, 2017},
  pages        = {1--18},
  publisher    = {{ACM}},
  year         = {2017},
  url          = {https://doi.org/10.1145/3132747.3132785},
  doi          = {10.1145/3132747.3132785},
  bibsource    = {dblp computer science bibliography, https://dblp.org}
}

@article{DBLP:journals/corr/abs-1905-05786,
  author       = {Joymallya Chakraborty and
                  Tianpei Xia and
                  Fahmid M. Fahid and
                  Tim Menzies},
  title        = {Software Engineering for Fairness: {A} Case Study with Hyperparameter
                  Optimization},
  journal      = {CoRR},
  volume       = {abs/1905.05786},
  year         = {2019},
  url          = {http://arxiv.org/abs/1905.05786},
  eprinttype    = {arXiv},
  eprint       = {1905.05786},
  bibsource    = {dblp computer science bibliography, https://dblp.org}
}

@article{DBLP:journals/software/Menzies20,
  author       = {Tim Menzies},
  title        = {The Five Laws of {SE} for {AI}},
  journal      = {{IEEE} Softw.},
  volume       = {37},
  number       = {1},
  pages        = {81--85},
  year         = {2020},
  url          = {https://doi.org/10.1109/MS.2019.2954841},
  doi          = {10.1109/MS.2019.2954841},
  bibsource    = {dblp computer science bibliography, https://dblp.org}
}

@inproceedings{felderer2021quality,
  title={Quality assurance for AI-based systems: Overview and challenges (introduction to interactive session)},
  author={Felderer, Michael and Ramler, Rudolf},
  booktitle={Software Quality: Future Perspectives on Software Engineering Quality: 13th International Conference, SWQD 2021, Vienna, Austria, January 19--21, 2021, Proceedings 13},
  pages={33--42},
  year={2021},
  organization={Springer}
}

@article{ma2018secure,
  title={Secure deep learning engineering: A software quality assurance perspective},
  author={Ma, Lei and Juefei-Xu, Felix and Xue, Minhui and Hu, Qiang and Chen, Sen and Li, Bo and Liu, Yang and Zhao, Jianjun and Yin, Jianxiong and See, Simon},
  journal={arXiv preprint arXiv:1810.04538},
  year={2018}
}

@inproceedings{shen2020multiple,
  title={Multiple-boundary clustering and prioritization to promote neural network retraining},
  author={Shen, Weijun and Li, Yanhui and Chen, Lin and Han, Yuanlei and Zhou, Yuming and Xu, Baowen},
  booktitle={Proceedings of the 35th IEEE/ACM International Conference on Automated Software Engineering},
  pages={410--422},
  year={2020}
}

@article{zhao2021state,
  title={State and tendency: an empirical study of deep learning question\&answer topics on Stack Overflow},
  author={Zhao, Henghui and Li, Yanhui and Liu, Fanwei and Xie, Xiaoyuan and Chen, Lin},
  journal={Science China Information Sciences},
  volume={64},
  pages={1--23},
  year={2021},
  publisher={Springer}
}

@inproceedings{islam2019comprehensive,
  title={A comprehensive study on deep learning bug characteristics},
  author={Islam, Md Johirul and Nguyen, Giang and Pan, Rangeet and Rajan, Hridesh},
  booktitle={Proceedings of the 2019 27th ACM Joint Meeting on European Software Engineering Conference and Symposium on the Foundations of Software Engineering},
  pages={510--520},
  year={2019}
}

@article{labib2022analysis,
  title={Analysis of noise and bias errors in intelligence information systems},
  author={Labib, Ashraf and Chakhar, Salem and Hope, Lorraine and Shimell, John and Malinowski, Mark},
  journal={Journal of the Association for Information Science and Technology},
  volume={73},
  number={12},
  pages={1755--1775},
  year={2022},
  publisher={Wiley Online Library}
}

@article{zhang2020machine,
  title={Machine learning testing: Survey, landscapes and horizons},
  author={Zhang, Jie M and Harman, Mark and Ma, Lei and Liu, Yang},
  journal={IEEE Transactions on Software Engineering},
  volume={48},
  number={1},
  pages={1--36},
  year={2020},
  publisher={IEEE}
}

@article{lan2022alphazero,
  title={Are AlphaZero-like Agents Robust to Adversarial Perturbations?},
  author={Lan, Li-Cheng and Zhang, Huan and Wu, Ti-Rong and Tsai, Meng-Yu and Wu, I and Hsieh, Cho-Jui and others},
  journal={Advances in Neural Information Processing Systems},
  volume={35},
  pages={11229--11240},
  year={2022}
}

@article{silver2017mastering,
  title={Mastering the game of go without human knowledge},
  author={Silver, David and Schrittwieser, Julian and Simonyan, Karen and Antonoglou, Ioannis and Huang, Aja and Guez, Arthur and Hubert, Thomas and Baker, Lucas and Lai, Matthew and Bolton, Adrian and others},
  journal={nature},
  volume={550},
  number={7676},
  pages={354--359},
  year={2017},
  publisher={Nature Publishing Group}
}

@article{liu2022deep,
  title={Deep learning and medical image analysis for COVID-19 diagnosis and prediction},
  author={Liu, Tianming and Siegel, Eliot and Shen, Dinggang},
  journal={Annual review of biomedical engineering},
  volume={24},
  pages={179--201},
  year={2022},
  publisher={Annual Reviews}
}

@article{chen2019looks,
  title={This looks like that: deep learning for interpretable image recognition},
  author={Chen, Chaofan and Li, Oscar and Tao, Daniel and Barnett, Alina and Rudin, Cynthia and Su, Jonathan K},
  journal={Advances in neural information processing systems},
  volume={32},
  year={2019}
}

@article{otter2020survey,
  title={A survey of the usages of deep learning for natural language processing},
  author={Otter, Daniel W and Medina, Julian R and Kalita, Jugal K},
  journal={IEEE transactions on neural networks and learning systems},
  volume={32},
  number={2},
  pages={604--624},
  year={2020},
  publisher={IEEE}
}

@article{lecun2015deep,
  title={Deep learning},
  author={LeCun, Yann and Bengio, Yoshua and Hinton, Geoffrey},
  journal={nature},
  volume={521},
  number={7553},
  pages={436--444},
  year={2015},
  publisher={Nature Publishing Group UK London}
}

@article{schuhmann2022laion,
  title={Laion-5b: An open large-scale dataset for training next generation image-text models},
  author={Schuhmann, Christoph and Beaumont, Romain and Vencu, Richard and Gordon, Cade and Wightman, Ross and Cherti, Mehdi and Coombes, Theo and Katta, Aarush and Mullis, Clayton and Wortsman, Mitchell and others},
  journal={Advances in Neural Information Processing Systems},
  volume={35},
  pages={25278--25294},
  year={2022}
}

@article{mittal2019survey,
  title={A survey of techniques for optimizing deep learning on GPUs},
  author={Mittal, Sparsh and Vaishay, Shraiysh},
  journal={Journal of Systems Architecture},
  volume={99},
  pages={101635},
  year={2019},
  publisher={Elsevier}
}

@misc{mnist-d,
    author="Y. LeCun and C. Cortes",
    title = "The MNIST database of handwritten digits",
    howpublished = "Website",
    year = {2019},
    note = {\url{http://yann.lecun.com/exdb/mnist/}},
    key = {"dataset1"}
}

@misc{cifar-10,
    author="N. Krizhevsky and H. Vinod and C. Geoffrey and M. Papadakis and A. Ventresque",
    title = "The cifar-10 dataset",
    howpublished = "Website",
    year = {2020},
    note = {\url{ http://www.cs.toronto.edu/~kriz/cifar.html}},
    key = {"dataset2"}
}

@article{Market,
  title={Person Re-identification Meets Image Search},
  author={Liang Zheng and Liyue Shen and Lu Tian and Shengjin Wang and Jiahao Bu and Qi Tian},
  journal={ArXiv},
  year={2015},
  volume={abs/1502.02171},
  url={https://api.semanticscholar.org/CorpusID:18597726}
}

@misc{resnet,
    title = "Resnet on the Cifar10 dataset",
    howpublished = "Website",
    year = {2022},
    note = {\url{https://github.com/akamaster/pytorch_resnet_cifar10}},
    key = {"model1"}
}

@misc{HRNet-18,
    title = "HRNet-18 on the Market dataset",
    howpublished = "Website",
    year = {2022},
    note = {\url{https://github.com/layumi/Person_reID_baseline_pytorch}},
    key = {"model5"}
}

@misc{siamese,
    title = "A Pytoch example of siamese network on mnist",
    howpublished = "Website",
    year = {2022},
    note = {\url{https://github.com/pytorch/examples/tree/main/siamese_network}},
    key = {"model3"}
}

@InProceedings{mff,
    author="Geoffrey Hinton",
    title="The Forward-Forward Algorithm: Some Preliminary Investigations",
    year="2022",
    note = {\url{https://arxiv.org/abs/2212.13345}},
    key = {"model4"}
}

@misc{mnist,
    title = "A Pytoch example of mnist",
    howpublished = "Website",
    year = {2022},
    note = {\url{https://github.com/pytorch/examples/tree/main/mnist}},
    key = {"model2"}
}

@misc{RAPL,
    author = "Srinivas Pandruvada",
    title = "RAPL Interface",
    howpublished = "Website",
    year = {2014},
    note = {\url{http://01.org/blogs/2014/running-average-power-limit-%E2%80%93-rapl}}
}

@misc{NVIDIA-SMI,
    title = "NVIDIA System Management Interface",
    howpublished = "Website",
    year = {2021},
    note = {\url{https://developer.nvidia.com/
nvidia- system- management- interface}},
    key = {"nvidia"}
}

@misc{perf,
    title = "perf: Linux profiling with performance counters",
    howpublished = "Website",
    year = {2020},
    note = {\url{https://perf.wiki.kernel. org/index.php/Main_Page}},
    key = {"perf"}
}

@inproceedings{RAPL-acc4, author = {Desrochers, Spencer and Paradis, Chad and Weaver, Vincent M.}, title = {A Validation of DRAM RAPL Power Measurements}, year = {2016}, isbn = {9781450343053}, publisher = {Association for Computing Machinery}, address = {New York, NY, USA}, url = {https://doi.org/10.1145/2989081.2989088}, doi = {10.1145/2989081.2989088}, booktitle = {Proceedings of the Second International Symposium on Memory Systems}, pages = {455–470}, numpages = {16}, location = {Alexandria, VA, USA}, series = {MEMSYS '16} }

@article{RAPL-acc3, author = {Khan, Kashif Nizam and Hirki, Mikael and Niemi, Tapio and Nurminen, Jukka K. and Ou, Zhonghong}, title = {RAPL in Action: Experiences in Using RAPL for Power Measurements}, year = {2018}, issue_date = {June 2018}, publisher = {Association for Computing Machinery}, address = {New York, NY, USA}, volume = {3}, number = {2}, issn = {2376-3639}, url = {https://doi.org/10.1145/3177754}, doi = {10.1145/3177754}, journal = {ACM Trans. Model. Perform. Eval. Comput. Syst.}, month = {mar}, articleno = {9}, numpages = {26} }

@article{RAPL-acc2,
author = {Kavanagh, Richard and Djemame, Karim},
title = {Rapid and accurate energy models through calibration with IPMI and RAPL},
journal = {Concurrency and Computation: Practice and Experience},
volume = {31},
number = {13},
pages = {e5124},
doi = {https://doi.org/10.1002/cpe.5124},
url = {https://onlinelibrary.wiley.com/doi/abs/10.1002/cpe.5124},
eprint = {https://onlinelibrary.wiley.com/doi/pdf/10.1002/cpe.5124},
note = {e5124 cpe.5124},
year = {2019}
}

@InProceedings{RAPL-acc1,
author="Paniego, Juan Manuel
and Gallo, Silvana
and Pi Puig, Mart{\'i}n
and Chichizola, Franco
and De Giusti, Laura
and Balladini, Javier",
editor="De Giusti, Armando Eduardo",
title="Analysis of RAPL Energy Prediction Accuracy in a Matrix Multiplication Application on Shared Memory",
booktitle="Computer Science -- CACIC 2017",
year="2018",
publisher="Springer International Publishing",
address="Cham",
pages="37--46",
isbn="978-3-319-75214-3"
}

@inproceedings{RAPL-use3, author = {Pereira, Rui and Couto, Marco and Saraiva, Jo\~{a}o and Cunha, J\'{a}come and Fernandes, Jo\~{a}o Paulo}, title = {The Influence of the Java Collection Framework on Overall Energy Consumption}, year = {2016}, isbn = {9781450341615}, publisher = {Association for Computing Machinery}, address = {New York, NY, USA}, url = {https://doi.org/10.1145/2896967.2896968}, doi = {10.1145/2896967.2896968}, booktitle = {Proceedings of the 5th International Workshop on Green and Sustainable Software}, pages = {15–21}, numpages = {7}, location = {Austin, Texas}, series = {GREENS '16} }

@INPROCEEDINGS{RAPL-use2,
  author={Moura, Irineu and Pinto, Gustavo and Ebert, Felipe and Castor, Fernando},
  booktitle={2015 IEEE/ACM 12th Working Conference on Mining Software Repositories}, 
  title={Mining Energy-Aware Commits}, 
  year={2015},
  volume={},
  number={},
  pages={56-67},
  doi={10.1109/MSR.2015.13}}

@INPROCEEDINGS{RAPL-use1,
  author={David, Howard and Gorbatov, Eugene and Hanebutte, Ulf R. and Khanna, Rahul and Le, Christian},
  booktitle={2010 ACM/IEEE International Symposium on Low-Power Electronics and Design (ISLPED)}, 
  title={RAPL: Memory power estimation and capping}, 
  year={2010},
  volume={},
  number={},
  pages={189-194},
  doi={10.1145/1840845.1840883}}

@inproceedings{greenAi, author = {Georgiou, Stefanos and Kechagia, Maria and Sharma, Tushar and Sarro, Federica and Zou, Ying}, title = {Green AI: Do Deep Learning Frameworks Have Different Costs?}, year = {2022}, isbn = {9781450392211}, publisher = {Association for Computing Machinery}, address = {New York, NY, USA}, url = {https://doi.org/10.1145/3510003.3510221}, doi = {10.1145/3510003.3510221}, booktitle = {Proceedings of the 44th International Conference on Software Engineering}, pages = {1082–1094}, numpages = {13}, location = {Pittsburgh, Pennsylvania}, series = {ICSE '22} }

@inproceedings{DeepFD, author = {Cao, Jialun and Li, Meiziniu and Chen, Xiao and Wen, Ming and Tian, Yongqiang and Wu, Bo and Cheung, Shing-Chi}, title = {DeepFD: Automated Fault Diagnosis and Localization for Deep Learning Programs}, year = {2022}, isbn = {9781450392211}, publisher = {Association for Computing Machinery}, address = {New York, NY, USA}, url = {https://doi.org/10.1145/3510003.3510099}, doi = {10.1145/3510003.3510099}, booktitle = {Proceedings of the 44th International Conference on Software Engineering}, pages = {573–585}, numpages = {13}, location = {Pittsburgh, Pennsylvania}, series = {ICSE '22} }

@INPROCEEDINGS{hrnet,
  author={Sun, Ke and Xiao, Bin and Liu, Dong and Wang, Jingdong},
  booktitle={2019 IEEE/CVF Conference on Computer Vision and Pattern Recognition (CVPR)}, 
  title={Deep High-Resolution Representation Learning for Human Pose Estimation}, 
  year={2019},
  volume={},
  number={},
  pages={5686-5696},
  doi={10.1109/CVPR.2019.00584}}

@INPROCEEDINGS{resnet18,
  author={He, Kaiming and Zhang, Xiangyu and Ren, Shaoqing and Sun, Jian},
  booktitle={2016 IEEE Conference on Computer Vision and Pattern Recognition (CVPR)}, 
  title={Deep Residual Learning for Image Recognition}, 
  year={2016},
  volume={},
  number={},
  pages={770-778},
  doi={10.1109/CVPR.2016.90}}

@article{gan, author = {Goodfellow, Ian and Pouget-Abadie, Jean and Mirza, Mehdi and Xu, Bing and Warde-Farley, David and Ozair, Sherjil and Courville, Aaron and Bengio, Yoshua}, title = {Generative Adversarial Networks}, year = {2020}, issue_date = {November 2020}, publisher = {Association for Computing Machinery}, address = {New York, NY, USA}, volume = {63}, number = {11}, issn = {0001-0782}, url = {https://doi.org/10.1145/3422622}, doi = {10.1145/3422622}, journal = {Commun. ACM}, month = {oct}, pages = {139–144}, numpages = {6} }

@ARTICLE{mutation_test,
  author={Jia, Yue and Harman, Mark},
  journal={IEEE Transactions on Software Engineering}, 
  title={An Analysis and Survey of the Development of Mutation Testing}, 
  year={2011},
  volume={37},
  number={5},
  pages={649-678},
  doi={10.1109/TSE.2010.62}}

@article{HONG,
title = {MUSEUM: Debugging real-world multilingual programs using mutation analysis},
journal = {Information and Software Technology},
volume = {82},
pages = {80-95},
year = {2017},
issn = {0950-5849},
doi = {https://doi.org/10.1016/j.infsof.2016.10.002},
url = {https://www.sciencedirect.com/science/article/pii/S0950584916302427},
author = {Shin Hong and Taehoon Kwak and Byeongcheol Lee and Yiru Jeon and Bongseok Ko and Yunho Kim and Moonzoo Kim}
}

@article{android, author = {Deng, Lin and Offutt, Jeff and Ammann, Paul and Mirzaei, Nariman}, title = {Mutation Operators for Testing Android Apps}, year = {2017}, issue_date = {January 2017}, publisher = {Butterworth-Heinemann}, address = {USA}, volume = {81}, number = {C}, issn = {0950-5849}, url = {https://doi.org/10.1016/j.infsof.2016.04.012}, doi = {10.1016/j.infsof.2016.04.012}, journal = {Inf. Softw. Technol.}, month = {jan}, pages = {154–168}, numpages = {15} }

@article{mem, author = {Wu, Fan and Nanavati, Jay and Harman, Mark and Jia, Yue and Krinke, Jens}, title = {Memory Mutation Testing}, year = {2017}, issue_date = {January 2017}, publisher = {Butterworth-Heinemann}, address = {USA}, volume = {81}, number = {C}, issn = {0950-5849}, url = {https://doi.org/10.1016/j.infsof.2016.03.002}, doi = {10.1016/j.infsof.2016.03.002}, journal = {Inf. Softw. Technol.}, month = {jan}, pages = {97–111}, numpages = {15} }

@article{aom, author = {Lindstr\"{o}m, Birgitta and Offutt, Jeff and Sundmark, Daniel and Andler, Sten F. and Pettersson, Paul}, title = {Using Mutation to Design Tests for Aspect-Oriented Models}, year = {2017}, issue_date = {January 2017}, publisher = {Butterworth-Heinemann}, address = {USA}, volume = {81}, number = {C}, issn = {0950-5849}, url = {https://doi.org/10.1016/j.infsof.2016.04.007}, doi = {10.1016/j.infsof.2016.04.007}, journal = {Inf. Softw. Technol.}, month = {jan}, pages = {112–130}, numpages = {19} }

@article{java-c,
title = {Source code optimization using equivalent mutants},
journal = {Information and Software Technology},
volume = {103},
pages = {138-141},
year = {2018},
issn = {0950-5849},
doi = {https://doi.org/10.1016/j.infsof.2018.06.013},
url = {https://www.sciencedirect.com/science/article/pii/S0950584918301332},
author = {Jorge López and Natalia Kushik and Nina Yevtushenko}
}

@incollection{industry,
title = {Chapter Six - Mutation Testing Advances: An Analysis and Survey},
editor = {Atif M. Memon},
series = {Advances in Computers},
publisher = {Elsevier},
volume = {112},
pages = {275-378},
year = {2019},
issn = {0065-2458},
doi = {https://doi.org/10.1016/bs.adcom.2018.03.015},
url = {https://www.sciencedirect.com/science/article/pii/S0065245818300305},
author = {Mike Papadakis and Marinos Kintis and Jie Zhang and Yue Jia and Yves Le Traon and Mark Harman}
}

@INPROCEEDINGS{api,
  author={Wen, Ming and Liu, Yepang and Wu, Rongxin and Xie, Xuan and Cheung, Shing-Chi and Su, Zhendong},
  booktitle={2019 IEEE/ACM 41st International Conference on Software Engineering (ICSE)}, 
  title={Exposing Library API Misuses Via Mutation Analysis}, 
  year={2019},
  volume={},
  number={},
  pages={866-877},
  doi={10.1109/ICSE.2019.00093}}

@INPROCEEDINGS{cpp,
  author={Álvarez-García and Miguel Ángel},
  booktitle={2021 IEEE/ACM 43rd International Conference on Software Engineering: Companion Proceedings (ICSE-Companion)}, 
  title={Automation and Evaluation of Mutation Testing for the New C++ Standards}, 
  year={2021},
  volume={},
  number={},
  pages={150-152},
  doi={10.1109/ICSE-Companion52605.2021.00063}}

@INPROCEEDINGS{cps,
  author={Cornejo, Oscar and Pastore, Fabrizio and Briand, Lionel},
  booktitle={2022 IEEE/ACM 44th International Conference on Software Engineering: Companion Proceedings (ICSE-Companion)}, 
  title={MASS: A tool for Mutation Analysis of Space CPS}, 
  year={2022},
  volume={},
  number={},
  pages={134-138},
  doi={10.1145/3510454.3516840}}

@INPROCEEDINGS{quantum,
  author={Fortunato, Daniel and Campos, José and Abreu, Rui},
  booktitle={2022 IEEE/ACM 44th International Conference on Software Engineering: Companion Proceedings (ICSE-Companion)}, 
  title={Mutation Testing of Quantum Programs Written in QISKit}, 
  year={2022},
  volume={},
  number={},
  pages={358-359},
  doi={10.1145/3510454.3528649}}

@inproceedings{jsimutate, author = {Laurent, Thomas and Arcaini, Paolo and Trubiani, Catia and Ventresque, Anthony}, title = {JSIMutate: Understanding Performance Results through Mutations}, year = {2022}, isbn = {9781450394130}, publisher = {Association for Computing Machinery}, address = {New York, NY, USA}, url = {https://doi.org/10.1145/3540250.3558930}, doi = {10.1145/3540250.3558930}, booktitle = {Proceedings of the 30th ACM Joint European Software Engineering Conference and Symposium on the Foundations of Software Engineering}, pages = {1721–1725}, numpages = {5}, location = {Singapore, Singapore}, series = {ESEC/FSE 2022} }

@INPROCEEDINGS{mutateion_tools,
  author={Ojdanic, Milos and Khanfir, Ahmed and Garg, Aayush and Degiovanni, Renzo and Papadakis, Mike and Le Traon, Yves},
  booktitle={2023 IEEE/ACM International Conference on Automation of Software Test (AST)}, 
  title={On Comparing Mutation Testing Tools through Learning-based Mutant Selection}, 
  year={2023},
  volume={},
  number={},
  pages={35-46},
  doi={10.1109/AST58925.2023.00008}}

@ARTICLE{pmt,
  author={Zhang, Jie and Zhang, Lingming and Harman, Mark and Hao, Dan and Jia, Yue and Zhang, Lu},
  journal={IEEE Transactions on Software Engineering}, 
  title={Predictive Mutation Testing}, 
  year={2019},
  volume={45},
  number={9},
  pages={898-918},
  doi={10.1109/TSE.2018.2809496}}

@article{Search-based,
title = {Search-based mutant selection for efficient test suite improvement: Evaluation and results},
journal = {Information and Software Technology},
volume = {104},
pages = {130-143},
year = {2018},
issn = {0950-5849},
doi = {https://doi.org/10.1016/j.infsof.2018.07.011},
url = {https://www.sciencedirect.com/science/article/pii/S0950584918301551},
author = {Pedro Delgado-Pérez and Inmaculada Medina-Bulo}
}

@inproceedings{inferring, author = {Just, Ren\'{e} and Kurtz, Bob and Ammann, Paul}, title = {Inferring Mutant Utility from Program Context}, year = {2017}, isbn = {9781450350761}, publisher = {Association for Computing Machinery}, address = {New York, NY, USA}, url = {https://doi.org/10.1145/3092703.3092732}, doi = {10.1145/3092703.3092732}, booktitle = {Proceedings of the 26th ACM SIGSOFT International Symposium on Software Testing and Analysis}, pages = {284–294}, numpages = {11}, location = {Santa Barbara, CA, USA}, series = {ISSTA 2017} }

@inproceedings{leam, author = {Tian, Zhao and Chen, Junjie and Zhu, Qihao and Yang, Junjie and Zhang, Lingming}, title = {Learning to Construct Better Mutation Faults}, year = {2023}, isbn = {9781450394758}, publisher = {Association for Computing Machinery}, address = {New York, NY, USA}, url = {https://doi.org/10.1145/3551349.3556949}, doi = {10.1145/3551349.3556949}, booktitle = {Proceedings of the 37th IEEE/ACM International Conference on Automated Software Engineering}, articleno = {64}, numpages = {13}, location = {Rochester, MI, USA}, series = {ASE '22} }

@INPROCEEDINGS{cnn-fiancail,
  author={Li, Da and Chen, Xinbo and Becchi, Michela and Zong, Ziliang},
  booktitle={2016 IEEE International Conferences on Big Data and Cloud Computing (BDCloud), Social Computing and Networking (SocialCom), Sustainable Computing and Communications (SustainCom) (BDCloud-SocialCom-SustainCom)}, 
  title={Evaluating the Energy Efficiency of Deep Convolutional Neural Networks on CPUs and GPUs}, 
  year={2016},
  volume={},
  number={},
  pages={477-484},
  doi={10.1109/BDCloud-SocialCom-SustainCom.2016.76}}

@inproceedings{dl-software, author = {Chen, Zhenpeng and Cao, Yanbin and Liu, Yuanqiang and Wang, Haoyu and Xie, Tao and Liu, Xuanzhe}, title = {A Comprehensive Study on Challenges in Deploying Deep Learning Based Software}, year = {2020}, isbn = {9781450370431}, publisher = {Association for Computing Machinery}, address = {New York, NY, USA}, url = {https://doi.org/10.1145/3368089.3409759}, doi = {10.1145/3368089.3409759}, booktitle = {Proceedings of the 28th ACM Joint Meeting on European Software Engineering Conference and Symposium on the Foundations of Software Engineering}, pages = {750–762}, numpages = {13}, location = {Virtual Event, USA}, series = {ESEC/FSE 2020} }

@article{GREEN, author = {Schwartz, Roy and Dodge, Jesse and Smith, Noah A. and Etzioni, Oren}, title = {Green AI}, year = {2020}, issue_date = {December 2020}, publisher = {Association for Computing Machinery}, address = {New York, NY, USA}, volume = {63}, number = {12}, issn = {0001-0782}, url = {https://doi.org/10.1145/3381831}, doi = {10.1145/3381831}, journal = {Commun. ACM}, month = {nov}, pages = {54–63}, numpages = {10} }

@INPROCEEDINGS{Estimating-Software-Energy,
  author={Fu, Cuijiao and Qian, Depei and Luan, Zhongzhi},
  booktitle={2018 IEEE International Conference on Internet of Things (iThings) and IEEE Green Computing and Communications (GreenCom) and IEEE Cyber, Physical and Social Computing (CPSCom) and IEEE Smart Data (SmartData)}, 
  title={Estimating Software Energy Consumption with Machine Learning Approach by Software Performance Feature}, 
  year={2018},
  volume={},
  number={},
  pages={490-496},
  doi={10.1109/Cybermatics_2018.2018.00106}}

@INPROCEEDINGS{Towards-a-Green,
  author={Mehra, Rohit and Sharma, Vibhu Saujanya and Kaulgud, Vikrant and Podder, Sanjay and Burden, Adam P.},
  booktitle={2022 IEEE/ACM 44th International Conference on Software Engineering: Software Engineering in Practice (ICSE-SEIP)}, 
  title={Towards a Green Quotient for Software Projects}, 
  year={2022},
  volume={},
  number={},
  pages={295-296},
  doi={10.1145/3510457.3513077}}

@INPROCEEDINGS{Optimising-workflow-execution,
  author={Warade, Mehul and Lee, Kevin and Ranaweera, Chathurika and Schneider, Jean-Guy},
  booktitle={2023 IEEE/ACM 7th International Workshop on Green And Sustainable Software (GREENS)}, 
  title={Optimising workflow execution for energy consumption and performance}, 
  year={2023},
  volume={},
  number={},
  pages={24-29},
  doi={10.1109/GREENS59328.2023.00011}}

@INPROCEEDINGS{Twins-or-False,
  author={Weber, Max and Kaltenecker, Christian and Sattler, Florian and Apel, Sven and Siegmund, Norbert},
  booktitle={2023 IEEE/ACM 45th International Conference on Software Engineering (ICSE)}, 
  title={Twins or False Friends? A Study on Energy Consumption and Performance of Configurable Software}, 
  year={2023},
  volume={},
  number={},
  pages={2098-2110},
  doi={10.1109/ICSE48619.2023.00177}}

@INPROCEEDINGS{Energy-Efficient-Machine,
  author={Kumar, Mohit and Zhang, Xingzhou and Liu, Liangkai and Wang, Yifan and Shi, Weisong},
  booktitle={2020 IEEE International Parallel and Distributed Processing Symposium Workshops (IPDPSW)}, 
  title={Energy-Efficient Machine Learning on the Edges}, 
  year={2020},
  volume={},
  number={},
  pages={912-921},
  doi={10.1109/IPDPSW50202.2020.00153}}

@article{What-can-Android, author = {Mcintosh, Andrea and Hassan, Safwat and Hindle, Abram}, title = {What Can Android Mobile App Developers Do about the Energy Consumption of Machine Learning?}, year = {2019}, issue_date = {April 2019}, publisher = {Kluwer Academic Publishers}, address = {USA}, volume = {24}, number = {2}, issn = {1382-3256}, url = {https://doi.org/10.1007/s10664-018-9629-2}, doi = {10.1007/s10664-018-9629-2}, journal = {Empirical Softw. Engg.}, month = {apr}, pages = {562–601}, numpages = {40} }

@INPROCEEDINGS{Black-Box-Technique,
  author={Bangash, Abdul Ali and Ali, Karim and Hindle, Abram},
  booktitle={2022 IEEE/ACM 44th International Conference on Software Engineering: New Ideas and Emerging Results (ICSE-NIER)}, 
  title={Black Box Technique to Reduce Energy Consumption of Android Apps}, 
  year={2022},
  volume={},
  number={},
  pages={1-5},
  doi={10.1145/3510455.3512795}}

@INPROCEEDINGS{Cost-effective-Strategies,
  author={Bangash, Abdul Ali},
  booktitle={2023 IEEE/ACM 45th International Conference on Software Engineering: Companion Proceedings (ICSE-Companion)}, 
  title={Cost-effective Strategies for Building Energy Efficient Mobile Applications}, 
  year={2023},
  volume={},
  number={},
  pages={281-285},
  doi={10.1109/ICSE-Companion58688.2023.00076}}

@article{pCAMP:-Performance-Comparison,
  author       = {Xingzhou Zhang and
                  Yifan Wang and
                  Weisong Shi},
  title        = {pCAMP: Performance Comparison of Machine Learning Packages on the
                  Edges},
  journal      = {CoRR},
  volume       = {abs/1906.01878},
  year         = {2019},
  url          = {http://arxiv.org/abs/1906.01878},
  eprinttype    = {arXiv},
  eprint       = {1906.01878},
  bibsource    = {dblp computer science bibliography, https://dblp.org}
}

@article{Estimation-of-energy, author = {Garc\'{\i}a-Mart\'{\i}n, Eva and Rodrigues, Crefeda Faviola and Riley, Graham and Grahn, H\r{a}kan}, title = {Estimation of Energy Consumption in Machine Learning}, year = {2019}, issue_date = {Dec 2019}, publisher = {Academic Press, Inc.}, address = {USA}, volume = {134}, number = {C}, issn = {0743-7315}, url = {https://doi.org/10.1016/j.jpdc.2019.07.007}, doi = {10.1016/j.jpdc.2019.07.007}, journal = {J. Parallel Distrib. Comput.}, month = {dec}, pages = {75–88}, numpages = {14} }

@INPROCEEDINGS{EREBA-Black-box,
  author={Haque, Mirazul and Yadlapalli, Yaswanth and Yang, Wei and Liu, Cong},
  booktitle={2022 IEEE/ACM 44th International Conference on Software Engineering (ICSE)}, 
  title={EREBA: Black-box Energy Testing of Adaptive Neural Networks}, 
  year={2022},
  volume={},
  number={},
  pages={835-846},
  doi={10.1145/3510003.3510088}}

@inproceedings{DeepXplore-Automated-Whitebox, author = {Pei, Kexin and Cao, Yinzhi and Yang, Junfeng and Jana, Suman}, title = {DeepXplore: Automated Whitebox Testing of Deep Learning Systems}, year = {2017}, isbn = {9781450350853}, publisher = {Association for Computing Machinery}, address = {New York, NY, USA}, url = {https://doi.org/10.1145/3132747.3132785}, doi = {10.1145/3132747.3132785}, booktitle = {Proceedings of the 26th Symposium on Operating Systems Principles}, pages = {1–18}, numpages = {18}, location = {Shanghai, China}, series = {SOSP '17} }

@INPROCEEDINGS{DeepGauge-Multi-Granularity,
  author={Ma, Lei and Juefei-Xu, Felix and Zhang, Fuyuan and Sun, Jiyuan and Xue, Minhui and Li, Bo and Chen, Chunyang and Su, Ting and Li, Li and Liu, Yang and Zhao, Jianjun and Wang, Yadong},
  booktitle={2018 33rd IEEE/ACM International Conference on Automated Software Engineering (ASE)}, 
  title={DeepGauge: Multi-Granularity Testing Criteria for Deep Learning Systems}, 
  year={2018},
  volume={},
  number={},
  pages={120-131},
  doi={10.1145/3238147.3238202}}

@inproceedings{MODE-automated-neural, author = {Ma, Shiqing and Liu, Yingqi and Lee, Wen-Chuan and Zhang, Xiangyu and Grama, Ananth}, title = {MODE: Automated Neural Network Model Debugging via State Differential Analysis and Input Selection}, year = {2018}, isbn = {9781450355735}, publisher = {Association for Computing Machinery}, address = {New York, NY, USA}, url = {https://doi.org/10.1145/3236024.3236082}, doi = {10.1145/3236024.3236082}, booktitle = {Proceedings of the 2018 26th ACM Joint Meeting on European Software Engineering Conference and Symposium on the Foundations of Software Engineering}, pages = {175–186}, numpages = {12}, location = {Lake Buena Vista, FL, USA}, series = {ESEC/FSE 2018} }

@INPROCEEDINGS{Guiding-Deep-Learning,
  author={Kim, Jinhan and Feldt, Robert and Yoo, Shin},
  booktitle={2019 IEEE/ACM 41st International Conference on Software Engineering (ICSE)}, 
  title={Guiding Deep Learning System Testing Using Surprise Adequacy}, 
  year={2019},
  volume={},
  number={},
  pages={1039-1049},
  doi={10.1109/ICSE.2019.00108}}

@INPROCEEDINGS{DeepCT-Tomographic-Combinatorial,
  author={Ma, Lei and Juefei-Xu, Felix and Xue, Minhui and Li, Bo and Li, Li and Liu, Yang and Zhao, Jianjun},
  booktitle={2019 IEEE 26th International Conference on Software Analysis, Evolution and Reengineering (SANER)}, 
  title={DeepCT: Tomographic Combinatorial Testing for Deep Learning Systems}, 
  year={2019},
  volume={},
  number={},
  pages={614-618},
  doi={10.1109/SANER.2019.8668044}}

@INPROCEEDINGS{Structural-Test-Coverage,
  author={Sun, Youcheng and Huang, Xiaowei and Kroening, Daniel and Sharp, James and Hill, Matthew and Ashmore, Rob},
  booktitle={2019 IEEE/ACM 41st International Conference on Software Engineering: Companion Proceedings (ICSE-Companion)}, 
  title={Structural Test Coverage Criteria for Deep Neural Networks}, 
  year={2019},
  volume={},
  number={},
  pages={320-321},
  doi={10.1109/ICSE-Companion.2019.00134}}

@INPROCEEDINGS{DeepLocalize-Fault-Localization,
  author={Wardat, Mohammad and Le, Wei and Rajan, Hridesh},
  booktitle={2021 IEEE/ACM 43rd International Conference on Software Engineering (ICSE)}, 
  title={DeepLocalize: Fault Localization for Deep Neural Networks}, 
  year={2021},
  volume={},
  number={},
  pages={251-262},
  doi={10.1109/ICSE43902.2021.00034}}

@INPROCEEDINGS{Apricot-A-Weight,
  author={Zhang, Hao and Chan, W.K.},
  booktitle={2019 34th IEEE/ACM International Conference on Automated Software Engineering (ASE)}, 
  title={Apricot: A Weight-Adaptation Approach to Fixing Deep Learning Models}, 
  year={2019},
  volume={},
  number={},
  pages={376-387},
  doi={10.1109/ASE.2019.00043}}

@INPROCEEDINGS{Importance-Driven-Deep,
  author={Gerasimou, Simos and Eniser, Hasan Ferit and Sen, Alper and Cakan, Alper},
  booktitle={2020 IEEE/ACM 42nd International Conference on Software Engineering: Companion Proceedings (ICSE-Companion)}, 
  title={Importance-Driven Deep Learning System Testing}, 
  year={2020},
  volume={},
  number={},
  pages={322-323},
  doi={}}

@INPROCEEDINGS{AUTOTRAINER-An-Automatic,
  author={Zhang, Xiaoyu and Zhai, Juan and Ma, Shiqing and Shen, Chao},
  booktitle={2021 IEEE/ACM 43rd International Conference on Software Engineering (ICSE)}, 
  title={AUTOTRAINER: An Automatic DNN Training Problem Detection and Repair System}, 
  year={2021},
  volume={},
  number={},
  pages={359-371},
  doi={10.1109/ICSE43902.2021.00043}}

@inproceedings{Amazon-SageMaker-Model, author = {Nigenda, David and Karnin, Zohar and Zafar, Muhammad Bilal and Ramesha, Raghu and Tan, Alan and Donini, Michele and Kenthapadi, Krishnaram}, title = {Amazon SageMaker Model Monitor: A System for Real-Time Insights into Deployed Machine Learning Models}, year = {2022}, isbn = {9781450393850}, publisher = {Association for Computing Machinery}, address = {New York, NY, USA}, url = {https://doi.org/10.1145/3534678.3539145}, doi = {10.1145/3534678.3539145}, booktitle = {Proceedings of the 28th ACM SIGKDD Conference on Knowledge Discovery and Data Mining}, pages = {3671–3681}, numpages = {11}, location = {Washington DC, USA}, series = {KDD '22} }

@inproceedings{UMLAUT-Debugging-Deep, author = {Schoop, Eldon and Huang, Forrest and Hartmann, Bjoern}, title = {UMLAUT: Debugging Deep Learning Programs Using Program Structure and Model Behavior}, year = {2021}, isbn = {9781450380966}, publisher = {Association for Computing Machinery}, address = {New York, NY, USA}, url = {https://doi.org/10.1145/3411764.3445538}, doi = {10.1145/3411764.3445538}, booktitle = {Proceedings of the 2021 CHI Conference on Human Factors in Computing Systems}, articleno = {310}, numpages = {16}, location = {Yokohama, Japan}, series = {CHI '21} }

@INPROCEEDINGS{Metamorphic-Testing-of,
  author={Gao, Wentao and He, Jiayuan and Pham, Van-Thuan},
  booktitle={2023 IEEE/ACM International Workshop on Deep Learning for Testing and Testing for Deep Learning (DeepTest)}, 
  title={Metamorphic Testing of Machine Translation Models using Back Translation}, 
  year={2023},
  volume={},
  number={},
  pages={1-8},
  doi={10.1109/DeepTest59248.2023.00008}}

@INPROCEEDINGS{Aries-Efficient-Testing,
  author={Hu, Qiang and Guo, Yuejun and Xie, Xiaofei and Cordy, Maxime and Papadakis, Mike and Ma, Lei and Traon, Yves Le},
  booktitle={2023 IEEE/ACM 45th International Conference on Software Engineering (ICSE)}, 
  title={Aries: Efficient Testing of Deep Neural Networks via Labeling-Free Accuracy Estimation}, 
  year={2023},
  volume={},
  number={},
  pages={1776-1787},
  doi={10.1109/ICSE48619.2023.00152}}

@INPROCEEDINGS{Information-Theoretic-Testing,
  author={Monjezi, Verya and Trivedi, Ashutosh and Tan, Gang and Tizpaz-Niari, Saeid},
  booktitle={2023 IEEE/ACM 45th International Conference on Software Engineering (ICSE)}, 
  title={Information-Theoretic Testing and Debugging of Fairness Defects in Deep Neural Networks}, 
  year={2023},
  volume={},
  number={},
  pages={1571-1582},
  doi={10.1109/ICSE48619.2023.00136}}

@INPROCEEDINGS{On-the-Cost,
  author={Arrieta, Aitor},
  booktitle={2022 IEEE/ACM 7th International Workshop on Metamorphic Testing (MET)}, 
  title={On the Cost-Effectiveness of Composite Metamorphic Relations for Testing Deep Learning Systems}, 
  year={2022},
  volume={},
  number={},
  pages={42-47},
  doi={10.1145/3524846.3527335}}

@INPROCEEDINGS{Adaptive-Test-Selection,
  author={Gao, Xinyu and Feng, Yang and Yin, Yining and Liu, Zixi and Chen, Zhenyu and Xu, Baowen},
  booktitle={2022 IEEE/ACM 44th International Conference on Software Engineering (ICSE)}, 
  title={Adaptive Test Selection for Deep Neural Networks}, 
  year={2022},
  volume={},
  number={},
  pages={73-85},
  doi={1O.1145/3510003.3510232}}

@inproceedings{Muffin-testing-deep, author = {Gu, Jiazhen and Luo, Xuchuan and Zhou, Yangfan and Wang, Xin}, title = {Muffin: Testing Deep Learning Libraries via Neural Architecture Fuzzing}, year = {2022}, isbn = {9781450392211}, publisher = {Association for Computing Machinery}, address = {New York, NY, USA}, url = {https://doi.org/10.1145/3510003.3510092}, doi = {10.1145/3510003.3510092}, booktitle = {Proceedings of the 44th International Conference on Software Engineering}, pages = {1418–1430}, numpages = {13}, location = {Pittsburgh, Pennsylvania}, series = {ICSE '22} }

@INPROCEEDINGS{EAGLE-Creating-Equivalent,
  author={Wang, Jiannan and Lutellier, Thibaud and Qian, Shangshu and Pham, Hung Viet and Tan, Lin},
  booktitle={2022 IEEE/ACM 44th International Conference on Software Engineering (ICSE)}, 
  title={EAGLE: Creating Equivalent Graphs to Test Deep Learning Libraries}, 
  year={2022},
  volume={},
  number={},
  pages={798-810},
  doi={10.1145/3510003.3510165}}

@article{Deep-Learning, author = {Heaton, Jeff}, title = {Ian Goodfellow, Yoshua Bengio, and Aaron Courville: Deep Learning}, year = {2018}, issue_date = {June 2018}, publisher = {Kluwer Academic Publishers}, address = {USA}, volume = {19}, number = {1–2}, issn = {1389-2576}, url = {https://doi.org/10.1007/s10710-017-9314-z}, doi = {10.1007/s10710-017-9314-z}, journal = {Genetic Programming and Evolvable Machines}, month = {jun}, pages = {305–307}, numpages = {3} }

@inproceedings{BLEU-a-method, author = {Papineni, Kishore and Roukos, Salim and Ward, Todd and Zhu, Wei-Jing}, title = {BLEU: A Method for Automatic Evaluation of Machine Translation}, year = {2002}, publisher = {Association for Computational Linguistics}, address = {USA}, url = {https://doi.org/10.3115/1073083.1073135}, doi = {10.3115/1073083.1073135}, booktitle = {Proceedings of the 40th Annual Meeting on Association for Computational Linguistics}, pages = {311–318}, numpages = {8}, location = {Philadelphia, Pennsylvania}, series = {ACL '02} }

@INPROCEEDINGS{Search-Based-Software,
  author={Sarro, Federica},
  booktitle={2023 IEEE 31st International Requirements Engineering Conference (RE)}, 
  title={Search-Based Software Engineering in the Era of Modern Software Systems}, 
  year={2023},
  volume={},
  number={},
  pages={3-5},
  doi={10.1109/RE57278.2023.00010}}

@article{Connecting-historical-changes, author = {Bai, Xue and Zhou, Hua and Yang, Hongji and Wang, Dong}, title = {Connecting Historical Changes for Cross-Version Software Defect Prediction}, year = {2020}, issue_date = {2020}, publisher = {Inderscience Publishers}, address = {Geneva 15, CHE}, volume = {63}, number = {4}, issn = {0952-8091}, url = {https://doi.org/10.1504/ijcat.2020.110428}, doi = {10.1504/ijcat.2020.110428}, journal = {Int. J. Comput. Appl. Technol.}, month = {jan}, pages = {371–383}, numpages = {12} }

@article{Cliffs-Delta-Calculator,
  title={Cliff's Delta Calculator: A non-parametric effect size program for two groups of observations},
  author={Macbeth, Guillermo and Razumiejczyk, Eugenia and Ledesma, Rub{\'e}n Daniel},
  journal={Universitas Psychologica},
  volume={10},
  number={2},
  pages={545--555},
  year={2011},
  publisher={Pontificia Universidad Javeriana}
}

@article{Comparison-of-values,
  title={Comparison of values of Pearson's and Spearman's correlation coefficients on the same sets of data},
  author={Hauke, Jan and Kossowski, Tomasz},
  journal={Quaestiones geographicae},
  volume={30},
  number={2},
  pages={87--93},
  year={2011}
}

\end{document}